\documentclass{article}

\usepackage[preprint]{neurips_2024}
\usepackage{natbib}

\usepackage{fancyhdr}

\usepackage{amsmath}
\usepackage{amsthm}
\usepackage{amssymb}
\usepackage{graphicx}
\usepackage{accents}
\usepackage{hyperref}
\usepackage{bbm}
\usepackage{bm}

\usepackage{float}
\usepackage{scalerel}
\usepackage{tikz,amsmath,siunitx}
\usetikzlibrary{decorations.pathreplacing}
\usetikzlibrary{shapes}

\usepackage{fancybox}

\usetikzlibrary{positioning}
\usetikzlibrary{shapes.multipart,chains,}

\newcommand{\RR}{\mathbb{R}}

\newcommand{\paren}[1]{\left( #1 \right)}
\newcommand{\br}[1]{\left[ #1 \right]}

\newcommand{\norm}[1]{\left\lVert#1\right\rVert}

\newcommand{\scr}[1]{\mathcal{#1}}

\newcommand{\noiter}{\#\textrm{it}}

\DeclareMathOperator*{\TT}{\operatorname{TT}}

\newtheorem{theorem}{Theorem}[section]

\newtheorem{lemma}[theorem]{Lemma}
\newtheorem{corollary}[theorem]{Corollary}
\newtheorem{definition}[theorem]{Definition}

\newcommand\myeq{\mathrel{\overset{\operatorname{def}}{=}}}
\usepackage{tikz,amsmath,siunitx}
\usetikzlibrary{decorations.pathreplacing}
\usetikzlibrary{shapes.geometric}
\usepackage{pifont}% http://ctan.org/pkg/pifont

\newcommand{\ts}{\mathsf{T}}
\usepackage{titlesec}
\titleformat*{\subsubsection}{\normalfont}
\usepackage{multirow}
\usepackage{wrapfig}
\usepackage{etoolbox}
\usepackage{float}
\usepackage[T1]{fontenc}    % use 8-bit T1 fonts
\usepackage{hyperref}       % hyperlinks
\usepackage{url}            % simple URL typesetting
\usepackage{booktabs}       % professional-quality tables
\usepackage{amsfonts}       % blackboard math symbols
\usepackage{nicefrac}       % compact symbols for 1/2, etc.
\usepackage{microtype}      % microtypography
\usepackage{xcolor}         % colors
\usepackage{algorithm}
\usepackage[noend]{algorithmic}
\usepackage{booktabs}
\usepackage{multirow}
\usepackage{caption}
\usepackage{subcaption}
\title{Efficient Leverage Score Sampling for Tensor Train Decomposition}

\newcommand\Mark[1]{\textsuperscript{#1}}

\author{
Vivek Bharadwaj\Mark{$^{*,1}$}, Beheshteh T. Rakhshan\Mark{$^{*,3}$}, Osman Asif Malik\Mark{$^2$}, Guillaume Rabusseau\Mark{$^{3,4}$}\\
\Mark{1}Electrical Engineering and Computer Science Department, UC Berkeley\\
  \Mark{2}Computational Research Division, Lawrence Berkeley National Lab\\
  \Mark{3}Mila \& DIRO,
Université de Montréal\\
\Mark{4}CIFAR AI Chair
}

\usepackage{lipsum}
\newif\ifAppendixlink

\Appendixlinktrue
\begin{document}
\maketitle
\def\thefootnote{*}\footnotetext{Equal contributions}
\begin{abstract}
Tensor Train~(TT) decomposition is widely used in the machine learning and quantum physics communities as a popular tool to efficiently compress high-dimensional tensor data. In this paper, we propose an efficient algorithm to accelerate computing the TT decomposition with the Alternating Least Squares (ALS) algorithm relying on exact leverage scores sampling. For this purpose, we propose a data structure that allows us to efficiently sample from the tensor with time complexity logarithmic in the tensor size. Our contribution specifically leverages the canonical form of the TT decomposition. By maintaining the canonical form through each iteration of ALS, we can efficiently compute (and sample from) the leverage scores, thus achieving significant speed-up in solving each sketched least-square problem. Experiments on synthetic and real data on dense and sparse tensors demonstrate that our method outperforms SVD-based and ALS-based algorithms. 

\end{abstract}

\section{Introduction}
Tensor decomposition methods have recently found numerous applications in machine learning. Their ability to perform operations efficiently on very high-dimensional tensors makes them suitable for data science and machine learning problems. For example, they have been used for neuro-imaging, and signal processing~\citep{Zhou2013,sidiropoulos2017tensor,cichocki2009fast}, supervised learning~\citep{novikov2016exponential,stoudenmire2016supervised}, feature extraction~\citep{bengua2015optimal} and scaling up Gaussian processes~\citep{izmailov2018scalable}. 
The two most popular decompositions are the CANDECOMP/PARAFAC~(CP) and Tucker decompositions~\citep{hitchcock1927expression,tucker1966some}. However, the number of parameters in the Tucker decomposition grows exponentially with the order of a tensor and finding a rank-$R$ CP decomposition is an NP-hard problem~\citep{kolda2009tensor,hillar2013most}. To address these issues, the Tensor Train~(TT) decomposition~\citep{oseledets2011tensor} can be used to represent a tensor in a compressed format where the number of parameters scales linearly with the order of a tensor. Also, there are stable algorithms to compute the TT decomposition.

Due to the high-dimensional nature of tensors, designing efficient algorithms for computing the TT decomposition is crucial.
A popular method for computing the TT decomposition of an $N$-dimensional tensor $\scr X$ is the TT-SVD algorithm~\citep{oseledets2011tensor} which uses a sequence of singular values decompositions on
the tensor unfoldings to produce the TT representation in a single pass. Since TT-SVD requires performing SVDs of unfoldings of $\scr X$, its cost is exponential in $N$. Alternating Least Square~(ALS) is another popular approach~\citep{holtz2012alternating} to find the TT approximation. Starting with a crude guess, each iteration of ALS involves solving a sequence of least squares problems. While ALS is the workhorse algorithm in many tensor decomposition problems, the computational cost is still exponential in $N$, since each iteration requires solving least squares problems involving unfoldings of $\scr X$. These issues have led to the search for alternatives based on randomization and sampling techniques. A cheaper alternative to the TT-SVD with strong accuracy guarantees can be implemented
by replacing the exact singular value decomposition~(SVD) with a well-studied randomized counterpart~\citep{halko2011finding,huber2017randomized}. Randomized variants of the TT-ALS approach have received little attention. In this work, we propose a novel randomized variant of the TT-ALS algorithm relying on exact leverage score sampling.

\textbf{Our Contributions.} In this paper, we propose a new sampling-based ALS approach to compute the TT decomposition: rTT-ALS. By using exact leverage score sampling, we are able to significantly reduce the size of each ALS least squares problem while providing strong guarantees on the approximation error. At the core of rTT-ALS, we leverage the TT canonical form to efficiently compute the exact leverage scores and speed up the solutions of least square problems in each iteration of ALS. To the best of our knowledge, rTT-ALS is the first efficient TT decomposition by the ALS algorithm which relies on leverage scores sampling. We provide experiments on synthetic and real massive sparse and dense tensors showing that rTT-ALS can achieve up to 26$\times$ speed-up compared to its non-randomized counterpart with little to no loss in accuracy. 
% \begin{itemize}
%     \item Propose a data structure to randomly select the rows of a design matrix from the exact distribution of its leverage scores to speed up the solution of the least-squares problems in each iteration of the ALS algorithm.
%     \item The leverage scores are computed efficiently by maintaining the canonical form of the TT decomposition in each iteration of the ALS algorithm. 
%     \item Taking advantage of the TT canonical form, we also obtain a faster convergence rate for solving the least-squares problems using the Preconditioned Conjugate Gradient method.
    % \item Demonstrate the effectiveness and efficiency of our proposed algorithm with the TT-cross method on synthetic and real-world data.
% \end{itemize}

Our core contribution is the following theorem, which shows that we can efficiently compute a subspace embedding of a left-orthogonal chain of TT tensor cores by efficiently sampling
according to their squared row norms.
\begin{theorem}[Row-norm-squared sampling for 3D core 
chains] Let $\scr A_1, ..., \scr A_j$ be a sequence of 3D tensors,
$\scr A_k \in \RR^{R_{k-1} \times I_k \times R_k}$~(with $R_0=1$). Assume that
the \textit{left-matricization} of each core is orthogonal. Let $A_{\leq j}$ be the $\prod_{k=1}^j I_k \times R_k$ matrix obtained by unfolding the contraction of the tensor chain $\scr A_1, ..., \scr A_j$. 
% defined by
% $$
% A_{\leq j}(\underline{i_1 \dots i_j}, r_j)= \sum_{r_0, ..., r_{j-1}} 
% \prod_{k=1}^j \scr{A}_k(r_{k-1}, i_k, r_k)
% $$
Then there exists a data structure to randomly sample rows
from $A_{\leq j}$ according to the distribution of its squared row norms
with the following properties:

\begin{enumerate}
    \item The data structure has construction time 
    $O \paren{\sum_{n=1}^j I_n R_{n-1} R_n^2}$. When 
    $R = R_1 = ... = R_j$ and $I = I_1 = ... = I_j$, the 
    runtime is $O(IR^3)$. The space overhead
    of the data structure is linear in the
    sizes of the input cores.

    \item The data structure produces a single
    row sample from $A_{\leq j}$ according to 
    the distribution of its squared row norms 
    in time $O \paren{ \sum_{k=1}^j \log \paren{I_k R_{k-1}/ R_{k}} R_k^2}$. When all ranks $R_k$ and physical dimensions
    $I_k$ are equal, this complexity is $O(j R^2 \log I)$. 
\end{enumerate}
\label{thm:main_result}
\end{theorem}
We highlight that the runtime required to construct the data structure is asymptotically identical to the runtime required to compute the canonical form of the tensor train subchain, i.e., $A_{\leq j}$, by successive QR decompositions.
%
% First, the 
% runtime required to construct the data structure is 
% asymptotically identical to the runtime required to compute
% a QR decomposition of each core. Second, when all ranks and
% physical dimensions are equal, the runtime required to draw
% a sample is only a factor $\log I$ worse than the cost to
% materialize a row of $A$ by tensor contraction. . Second, when all ranks and
% physical dimensions are equal, the runtime required to draw
% a sample is only a factor $\log I$ worse than the cost to
% materialize a row of $A$ by tensor contraction. 
% We highlight two facts about this theorem. First, the 
% runtime required to construct the data structure is 
% asymptotically identical to the runtime required to compute
% a QR decomposition of each core. Second, when all ranks and
% physical dimensions are equal, the runtime required to draw
% a sample is only a factor $\log I$ worse than the cost to
% materialize a row of $A$ by tensor contraction. 

% I'm putting the proof in section 4 

\section{Related work}
Randomized algorithms and leverage score sampling-based methods~\citep{mahoney2011randomized,woodruff2014sketching,drineas2006fast} have been used widely in a large body of research particularly in tensor decomposition problems over the past two decades~\citep{malik2021sampling,bharadwaj2023fast,larsen2022practical,fahrbach2022subquadratic} just to name a few.

\citep{cheng2016spals} propose SPALS, the first ALS-based algorithm relying on leverage score sampling for the CP decomposition. Their proposed method reduces the size of the least squares problem in each iteration of ALS with a sub-linear cost per iteration in the number of entries of the input tensor. ~\citet{larsen2022practical} extends this method by combining repeated sampled rows in a deterministic and random sampling fashion. 
However, both of these methods use leverage score approximations and therefore require a number of samples which is exponential in the number of tensor modes in order to achieve relative-error performance guarantees. 
\citet{malik2022more} proposes a method which 
avoids this exponential dependency on the number of tensor modes by using higher-quality leverage score estimates for the CP decomposition.
The method is further improved by 
\citep{malik2022samplingbased} to use exact rather than approximate leverage scores which is applicable for arbitrary tensor decompositions.
Recently,~\citep{bharadwaj2023fast} provided a novel data structure to efficiently sample from the exact distribution of the factor matrices'
leverage scores in the Khatri-Rao product with time complexity logarithmic in the tensor size, leading to further improvements on the work in \citep{malik2022samplingbased}. The sampler we propose in this paper is built on the work by \cite{bharadwaj2023fast}, extending it to the TT decomposition and leveraging the canonical form for further speed-up. 
%However, it differs from the aforementioned papers since they are all concerned with the CP decomposition and our paper considers randomized ALS-based algorithm for 
%the TT decomposition. 

There are also a variety of non-ALS-based randomized algorithms for computing the TT decomposition.~\citep{huber2017randomized} leverages randomized SVD for the TT decomposition which accelerates the classical TT-SVD algorithm proposed by~\citep{oseledets2011tensor}.
To handle situations where the exact TT rank is unknown, ~\citep{che2019randomized} propose an adaptive randomized algorithm which is able to obtain a nearly optimal TT approximation.
~\citep{yu2023randomized} present an algorithm for computing the TT approximation using randomized block Krylov subspace iteration. 
More precisely, most of the randomized algorithms for the TT decomposition are based on the randomized SVD for matrices which is introduced by~\citep{halko2011finding}. More closely related to our work are those using sketching and sampling in each iteration of ALS algorithm to approximate the TT decomposition. Recently,~\citep{chen2023low} leverage TensorSketch~\citep{pham2013fast} in each iteration of a regularized ALS approach for TT decomposition.

\section{Preliminaries}
We use capital letters $A$ to denote matrices and script characters $\scr A$ to denote multidimensional arrays. We use Matlab notation for slices of matrices and tensors. We use the tuple notation to indicate the position of entries of arrays. For example, $\scr A(i_1,i_2,i_3)$ indicates the $(i_1,i_2,i_3)$-th element of $\scr A$. 
$A\br{i, :}$ and $A\br{:, i}$ refer to the $i$-th row and column of $A$, respectively; for a three-dimensional tensor $\scr A \in \RR^{R_1 \times I_1 \times R_2}$, the matrix $\scr A \br{:, i, :} \in \RR^{R_1 \times R_2}$ is the $i$-th lateral slice of $\scr A$. For a positive integer $n$, we use $[n]$ to denote the set of integers from 1 to $n$. For $i_1\in [I_1], \dots, i_N\in [I_N]$, the notation $\underline{i_1 \dots i_n}\myeq 1 + \sum_{n =1}^N (i_n-1)\prod_{j=1}^{N-1}I_j$ will be helpful for tensor unfoldings. We use $\otimes$ and $\odot$ to denote the Kronecker and Khatri-Rao products, respectively~(see definitions in Appendix~\ref{appendix:additinal-notations}). We use $I_d$ to denote the $d\times d$ identity matrix, $A^{\ts}$ for the transpose of  $A$ , $A^+$ for the pseudo-inverse of $A$, $\norm{\cdot}_F$ for the Frobenius norm and $\norm{\cdot}_2$ for the Euclidean norm of a vector. We use $\tilde{O}$ to indicate multiplicative terms polylogarithmic in $R$ and $1/\delta$.
\subsection{Tensor Train Decomposition}
Let $\scr X\in\RR^{I_1\times \cdots \times I_N}$ be an $N$-dimensional array.  A rank $(R_1,\dots,R_{N-1})$ \textit{tensor train (TT) decomposition} of a tensor $\scr X\in\RR^{I_1\times\cdots\times I_N}$ factorizes it into the product of $N$  third-order tensors  $\scr A_n\in\RR^{R_{n-1}\times I_n\times R_n}$ for $n\in [N]$~(with $R_0=R_N=1$):
$$\scr X(i_1,\cdots,i_N)= \sum_{r_0,\cdots, r_N}\prod_{n=1}^N\scr A_n(r_{n-1},i_n,r_n),$$ 
for all $i_1\in[I_1],\cdots,i_N\in[I_N]$, where each $r_n$ ranges from 1 to $R_n$.  A tensor network representation of a TT decomposition is shown in Figure~\ref{fig:tt-decomp}. We call $\scr A_1,\scr A_2,\cdots, \scr A_N$ core tensors and we use $\TT(({\scr A_n})_{n=1}^N)$ to denote a TT tensor with factors $\scr A_1, \cdots, \scr A_n$.
% Throughout this work, we consider tensor trains with sufficiently low rank (below 100). 
% Matrix product states~\cite{oseledets2010approximation} that arise in quantum physics frequently do not satisfy this
% property, with ranks up to several hundred that significantly exceed the tensor 
% train physical dimensions. On the other hand, these objects typically have additional block 
% structure within the cores that require more tailored algorithms than the general 
% methods that we present here.
\begin{figure}[h!]
    \centering
\begin{tikzpicture}[baseline=-0.5ex]
    \tikzset{tensor/.style = {minimum size = 0.5cm,shape = circle,thick,draw=black,inner sep = 0pt}, edge/.style = {thick,line width=.4mm},every loop/.style={}}
    \def\x{6}
    \def\y{0}
     \node[tensor,fill=green!50!red!50!white] (B) at (\x+1,\y) {$\scr X$};
    \draw[edge] (B) -- (\x+0.3,0) node [midway,above] {\scalebox{0.7}{\textcolor{gray}{$i_1$}}};
    \draw[edge] (B) -- (\x+1.6,0) node [midway,above] {\scalebox{0.7}{\textcolor{gray}{$i_5$}}};;
    \draw[edge] (B) -- (\x+0.4,-0.3) node [midway, below] {\scalebox{0.7}{\textcolor{gray}{$i_2$}}};;
    \draw[edge] (B) -- (\x+1.5,-0.4)node [midway, below] {\scalebox{0.7}{\textcolor{gray}{$i_4$}}};;
     \draw[edge] (B) -- (\x+1,-0.7)node [below] {\scalebox{0.7}{\textcolor{gray}{$i_3$}}};;
\end{tikzpicture}
=
\begin{tikzpicture}[baseline=-0.5ex]
    \tikzset{tensor/.style = {minimum size = 0.5cm,shape = circle,thick,draw=black,inner sep = 0pt}, edge/.style   = {thick,line width=.4mm},every loop/.style={}}
    \def\y{0}
    \def\x{0}
  \node[tensor,fill=red!30!white] (D) at (\x-1,\y){$\scr A_1$};
  \node[tensor,fill=red!50!white] (A) at (\x,\y){\scalebox{0.85}{$\scr A_2$}};
  \node[tensor,fill=blue!50!white] (B) at (\x+1,\y){$\scr A_3$};
   \node[tensor,fill = blue!60!red!50!] (C) at (\x+2,\y){$\scr A_4$};
   \node[tensor,fill = blue!20!white] (E) at (\x+3,\y){$\scr A_5$};
    \draw[edge](A) -- (\x,\y-0.7)node [midway,left] {\scalebox{0.7}{\textcolor{gray}{$i_2$}}};
    \draw[edge](B) -- (\x+1,\y-0.7)node [midway,left] {\scalebox{0.7}{\textcolor{gray}{$i_3$}}};
    \draw[edge](C) -- (\x+2,\y-0.7)node [midway,left] {\scalebox{0.7}{\textcolor{gray}{$i_4$}}};
     \draw[edge](D) -- (\x-1,\y-0.7)node [midway,left] {\scalebox{0.7}{\textcolor{gray}{$i_1$}}};
     \draw[edge](E) -- (\x+3,\y-0.7)node [midway,left] {\scalebox{0.7}{\textcolor{gray}{$i_5$}}};
    \draw[edge](D) -- (A) node [midway,above] {\scalebox{0.7}{\textcolor{gray}{$R_1$}}};
    \draw[edge](A) -- (B) node [midway,above] {\scalebox{0.7}{\textcolor{gray}{$R_2$}}};
    \draw[edge](B) -- (C) node [midway,above] {\scalebox{0.7}{\textcolor{gray}{$R_3$}}};
    \draw[edge](C) -- (E) node [midway,above] {\scalebox{0.7}{\textcolor{gray}{$R_4$}}};
\end{tikzpicture}
\caption{Tensor Train decomposition of a 5-dimensional tensor in tensor network notation.}\label{fig:tt-decomp}
\end{figure}
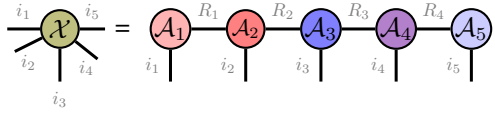
\begin{definition}\label{def:n-fold-matricization}
    The mode-$n$ unfolding of a tensor $\scr X\in\RR^{I_1\times\dots\times I_N}$ is the matrix $X_{(n)}\in\RR^{I_n\times\prod_{j\neq n}I_j}$ defined element-wise by
    $X_{(n)}\left(i_n,\underline{i_1\cdots i_{n-1}i_{n+1}\cdots i_N}\right)\myeq \scr X(i_1,\cdots,i_N).$

    As a special case, we denote the left~(resp. right) matricization of a 3-dimensional tensor $\scr A\in\RR^{I_1\times I_2 \times I_3}$ by $A^L = (A)_{(3)}^\top\in\RR^{I_1I_2\times I_3}$ and $A^R = A_{(1)}\in\RR^{I_1\times I_2I_3}$.
\end{definition}

Given a TT decomposition $\TT(({\scr A_n})_{n=1}^N)$ and an index $j$, we will often use the left-chain $A_{<j}\in\RR^{\prod_{k=1}^{j-1}I_k\times R_{j-1}}$ and right-chain ~$A_{>j}\in\RR^{R_j\times\prod_{k={j+1}}^NI_k}$ unfoldings obtained by matricizing the contraction of all cores on the left and on the right side of the $j$-th core. Formally,
$$
  A_{<j}\left(i_{<j},r_{j-1}\right) =\!\!\! \sum_{r_0,\dots,r_{j-1}}\prod_{k=1}^{j-1}\scr A_k(r_{k-1},i_k,r_k)\ \ \text{and}\ \ 
   A_{>j}\left(r_j,i_{>j}\right) =\!\!\!  \sum_{r_{j+1},\dots,r_{N}}\prod_{k=j+1}^{N}\scr A_k(r_{k-1},i_k,r_k)
$$
where $i_{<j}=\underline{i_1\dots i_{j-1}}$ and $i_{>j}=\underline{i_{j+1}\dots i_{N}}$.
We also use $A^{\neq j} \overset{\operatorname{def}}{=}	 A_{<j}\otimes A_{>j}^\top\in\mathbb{R}^{\prod_{k\neq j}I_k\times R_{j-1}R_j}$ to denote the unfolding of the contraction of all cores except the $j$-th one. 

\begin{figure}[h!]

    \centering
\begin{tikzpicture}[baseline=-0.5ex]
    \tikzset{tensor/.style = {minimum size = 0.5cm,shape = circle,thick,draw=black,inner sep = 0pt}, edge/.style = {thick,line width=.4mm},every loop/.style={}}
    \def\x{0}
    \def\y{0}
 \node[tensor, path picture={
        \fill[fill=red!30!white](path picture bounding box.south east)
        -- 
        (path picture bounding box.north west) --
        (path picture bounding box.north east) -- cycle;
        }, shift={(\x,\y)}] (A) {\scalebox{0.85}{$\scr A_1$}};
        \draw[edge](A) -- (\x,\y-0.7) node [midway,left] {\scalebox{0.7}{\textcolor{gray}{$i_1$}}};
       \node[tensor, path picture={
        \fill[fill=red!50!white](path picture bounding box.south east)
        -- 
        (path picture bounding box.north west) --
        (path picture bounding box.north east) -- cycle;
        }, shift={(\x+1,\y)}] (B) {\scalebox{0.85}{$\scr A_2$}};
         \draw[edge](B) -- (\x+1,\y-0.7)node [midway,left] {\scalebox{0.7}{\textcolor{gray}{$i_2$}}};
         \node[tensor,fill=blue!50!white] (C) at (\x+2,\y){$\scr A_3$};
          \draw[edge](C) -- (\x+2,\y-0.7)node [midway,left] {\scalebox{0.7}{\textcolor{gray}{$i_3$}}};
          \node[tensor, path picture={
        \fill[fill=blue!60!red!50!](path picture bounding box.south west)
        -- 
        (path picture bounding box.north east) --
        (path picture bounding box.north west) -- cycle;
        }, shift={(\x+3,\y)}] (D) {\scalebox{0.85}{$\scr A_4$}};
         \draw[edge](D) -- (\x+3,\y-0.7)node [midway,left] {\scalebox{0.7}{\textcolor{gray}{$i_4$}}};
           \node[tensor, path picture={
        \fill[fill=blue!20!white](path picture bounding box.south west)
        -- 
        (path picture bounding box.north east) --
        (path picture bounding box.north west) -- cycle;
        }, shift={(\x+4,\y)}] (E) {\scalebox{0.85}{$\scr A_5$}};
         \draw[edge](E) -- (\x+4,\y-0.7)node [midway,left] {\scalebox{0.7}{\textcolor{gray}{$i_5$}}};
    \draw[edge](A) -- (B) node [midway,above] {\scalebox{0.7}{\textcolor{gray}{$R_1$}}};
    \draw[edge](B) -- (C) node [midway,above] {\scalebox{0.7}{\textcolor{gray}{$R_2$}}};
    \draw[edge](C) -- (D) node [midway,above] {\scalebox{0.7}{\textcolor{gray}{$R_3$}}};
    \draw[edge](D) -- (E) node [midway,above] {\scalebox{0.7}{\textcolor{gray}{$R_4$}}};
\end{tikzpicture}
\caption{Orthonormal TT decomposition. The cores at the left side of $\scr A_3$ are left-orthonormal and the cores at the right are right-orthonormal.}\label{fig:orthonormal-tt}
\end{figure}
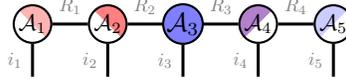

We conclude by introducing the canonical form of the TT decomposition~\citep{holtz2012alternating,evenbly2018gauge,evenbly2022practical} which will be central to the design of our algorithm.
\begin{definition}\label{def:orthonormal-cores}
   A TT decomposition $\TT(({\scr A_n})_{n=1}^N)\in\RR^{I_1\times\dots\times I_N}$ is in a canonical format with respect to a fixed index $j\in [N]$ if ${A_n^L}^\top A_n^L = I_{R_n}$  for all $n<j$, and $A_n^R{A_n^R}^\top = I_{R_{n-1}}$ for all $n>j$~(see Figure~\ref{fig:orthonormal-tt}).
\end{definition}
Note that any TT decomposition can efficiently be converted to canonical form w.r.t. any index $j\in [N]$ by performing a series of QR decompositions on the core tensors.
\subsection{Alternating Least Squares with Tensor Train Structure.}
%The goal is to fit a TT decomposition $\TT(({\scr A_n})_{n=1}^N)$ to a data tensor $\scr X$ such that the value $\norm{\scr X - \TT(({\scr A_n})_{n=1}^N)}_F$ is minimal. 
The TT decomposition problem consists in finding a low-rank approximation $\TT(({\scr A_n})_{n=1}^N)$ of a given  tensor $\scr X$:
$
    \text{argmin}_{\scr A_1,\dots,\scr A_N}\norm{\scr X - \TT(\scr A_1,\dots,\scr A_N)}_F  
$
where $\scr {X}$ is the target tensor with dimensions $I_1\times\dots\times I_N$. 
Since this is a non-convex optimization problem, the popular alternating least-squares~(ALS) approach can be used to find an approximate solution~\citep{kolda2009tensor}. Fixing all cores except the $j$-th one, the low rank approximation problem can be reformulated as a linear least squares problem: 
\begin{align}\label{eq:lstsq_task-als}
\text{argmin}_{\scr A_j}\norm{\paren{A_{<j} \otimes A_{>j}^\top} (A_j)_{(2)}^\top - X_{(j)}^\top}_F. 
\end{align}
The ALS approach finds an approximate solution by keeping all cores fixed and solving for the $j$-th one. Then repeat this procedure multiple times for each $j\in[N]$ until some convergence criteria is met. In this work, we will combine ALS with core orthogonalization to efficiently compute the exact leverage scores. This will also lead to a stable algorithm for computing TT. To compute the orthogonalized TT approximation, we start with a crude TT decomposition in canonical form~(see Definition~\ref{def:orthonormal-cores}) where all cores except the first one are right-orthonormal. After optimizing the first core, a QR decomposition is performed and the non-orthonormal part is merged into the next core. This procedure repeats until reaching the right side of the decomposition. The same procedure is then repeated from the right until reaching the left side~(see the tensor network illustration in Appendix~\ref{app:orthogonal-tt}). This approach leads to providing computational benefits for computing the leverage scores and to an efficient sampling scheme which will be discussed in Section~\ref{efficient-tt-sampling}. 
%
%
%
%
% \begin{algorithm}[H]
%     \caption{TT-ALS}\label{alg:ALS}
%     \hspace*{\algorithmicindent} \textbf{Input:}~$\scr X\in\RR^{I_1\times\dots\times I_N}$, target ranks $(R_1,\dots,R_N)$\\
%     \hspace*{\algorithmicindent} \textbf{Output:} TT cores~$\scr A_1,\dots,\scr A_N$
%     \begin{algorithmic}[1]
%     \STATE Initialize {$\scr A_2,\dots, \scr A_N $}
%     \WHILE{some termination criteria is met}
%     \FOR {$n=1,\dots,N,\dots,1$}
%         \STATE $\scr A_n = \text{argmin}_{\scr A}\norm{A^{\neq n}A_{(2)}^\top - X_{(n)}^\top}_F$\label{agl:line4}
%         \ENDFOR
%     \ENDWHILE
%     \STATE \textbf{return} $\scr A_1, \dots, \scr A_N$
%     \end{algorithmic}
%     \label{alg:chain_sampling}
% \end{algorithm}
%
%
%
%
\subsection{Sketching and Leverage Score Sampling}
There exists a vast literature on randomized algorithms~\citep{mahoney2011randomized,woodruff2014sketching} to solve the over-determined 
least squares problem $\text{min}_x\norm{Ax-b}_F$ where 
$A\in\RR^{I\times R}, I\gg R$. Regardless of the structure of both $A$ and 
$b$, solving this least-squares problem costs $O(I R^2)$. To reduce this cost, we can randomly select rows of $A$ and $b$ by proposing a sketching operator $S$ with $J\ll I$. 
Therefore, instead of solving the original least squares problem, we consider solving the downsampled version of the form $\min_x \norm{SAx- Sb}_F$, where $S\in\RR^{J\times I}$ and reduce the cost to $O(JR^2)$. The goal is to find a ``good'' sketch $S$ to approximate 
the solution of the least squares problem at each step of the ALS algorithm. When each entry of $S$ is selected according to the rows of $A$ leverage scores, strong guarantees can be obtained for the solution of the downsampled problem.  

\begin{definition}(Leverage scores)\label{def:leverage-scores}
    Suppose $A\in\RR^{I\times R}$ with $I\gg R$. The $i$-th leverage score of the matrix $A$ is defined as 
\begin{align}\label{eq:leverage-scores}
   l_i(A) = A[i,:](A^\top A)^+ A[i,:]^\top
   ~~\text{for}~~i\in[I].
\end{align}
%where $(A^\top A)$ is the so-called \textit{Gram} matrix.
\end{definition}
\begin{definition}(Leverage score sampling)\label{def:leverage-score sampling}
    Let $A\in\RR^{I\times R}$ and $p\in[0,1]^I$ be a probability distribution vector with entries $p_i = \frac{l_i(A)}{\text{rank}(A)}$; where $\text{rank}(A) = \sum_i l_i(A)$. Assume $\hat s_1, ..., \hat s_J$ are drawn i.i.d according to the probabilities $p_1,\cdots,p_I$. The random matrix $S\in\RR^{J\times I}$ defined element-wise by 
        $S(j,i) = 
        \frac{1}{\sqrt{Jp_i}}$
    if $\hat s_j =i$ 
and 0 otherwise is called a leverage score sampling matrix for $A$.
\end{definition}
The following result is well-known and appeared in several works; see, e.g., \citep{drineas2006SamplingAlgorithms}, \citep{drineas2008RelativeerrorCUR}, \citep{drineas2011faster}, \citep{larsen2022practical}. We borrow the form presented in~\citep{malik2022more}.
\begin{theorem}(Guarantees for Leverage Score Sampling)\label{thm:relative-error gaurantee}
Suppose $A\in\RR^{I\times R}$. Let $S\in\RR^{J\times I}$ be the leverage score sampling matrix defined in~\ref{def:leverage-score sampling}. For any $\varepsilon,\delta\in(0,1)$, if $J = \tilde{O} (R^2/\varepsilon\delta)$, then
$\tilde{x}^* = \min_x\norm{SAx -Sb}_2$ satisfies
    $\norm{Ax^*-b}_2
    \leq
    (1+\varepsilon)\min_x\norm{Ax-b}_2,$ with probability $1-\delta.$
\end{theorem}
Computing leverage scores in Definition~\ref{def:leverage-scores} requires computing the pseudo-inverse of $A$, which costs $O(IR^2)$ and is as costly as directly solving the original least squares problem. In the following section, we will show that the leverage scores can be computed much more efficiently when $A$ is the matrix appearing in the TT-ALS algorithm in canonical form..

\section{Sampling-based Tensor Train Decomposition}\label{efficient-tt-sampling}
In this section, we show how to efficiently sample rows of $A^{\neq j} = A_{<j} \otimes A_{>j}^\top$ and $X_{(j)}$ in Equation~\eqref{eq:lstsq_task-als} according to the exact leverage scores distribution. In doing so, we will also present the sketch of the proof of Theorem~\ref{thm:main_result}~(which closely mirrors
that of \cite{bharadwaj2023fast} with key modifications required to adapt the procedure to a tensor core chain).

%Since $A_{<j}$ and $A_{>j}$ are both orthonormal and the Kronecker product of them is also orthonormal, we propose a sampler that exploits the canonical form of $A^{\neq j}$. 
For each row $i^{\neq j} = \underline{i_1\dots i_{j-1}i_{j+1}\dots i_N}$ of $A^{\neq j}$, Equation~\eqref{eq:leverage-scores} gives
\begin{equation}\label{eq:lev_Anotj}
l_{i^{\neq j}}(A^{\neq j})= A^{\neq j}[i^{\neq j},:](A^{{\neq j}^{\top}} A^{\neq j})^+A^{\neq j}[:,i^{\neq j}]^\top.
\end{equation}
Computing $\Phi \myeq (A^{{\neq j}^{\top}} A^{\neq j})^+$ is the main computational bottleneck in finding the leverage scores of $A^{\neq j}$. \citet{malik2022samplingbased} proposed an algorithm to compute $\Phi$ in time $O(NIR^2 + R^3)$. In this paper, we leverage the fact that when the TT tensor is in canonical form w.r.t. mode $j$, $A^{\neq j}$ is orthogonal, and thus $\Phi = I_{R^2}$. Therefore, computing $\Phi$ is free of cost. By maintaining the canonical form of the TT tensor throughout the ALS algorithm, we can sketch the least square problems from the leverage score distributions with almost no computational overhead. 
We now explain how to efficiently sample rows of $A^{\neq j}$ from the leverage scores distribution.

\subsection{Efficient Core Chain Leverage Score Sampling}
% Each least squares problem in Equation~\eqref{eq:lstsq_task-als} contains all entries of $\scr X$. To reduce the size of each least squares problem, we propose a sampler that allows us to sample a row from the exact leverage scores. More importantly, it allows us to sample of $A^{\neq j}$. That means we can sample either of $A_{<j}$ or $A_{>j}^\top$ with probability proportional to
% its squared row norms~(due to the orthonormality property of $A_{<j}$ and $A_{>j}$). Without loss of generality and by using Kronecker product property and leveraging canonical form, we detail the sampling procedure of the $A_{\leq j}$~(the difference between $A_{\leq j}$ and $A_{<j}$ amounts to reindexing). The sampling procedure of $A_{>j}$ will be the same and straightforward.  
As discussed above, when the TT tensor is in canonical form, the leverage score of row $i^{\neq j}$ is given by $l_{i^{\neq j}}(A^{\neq j})= A^{\neq j}[i^{\neq j},:]A^{\neq j}[:,i^{\neq j}]^\top$. Leveraging the Kronecker structure of $A^{\neq j}= A_{<j} \otimes A_{>j}^\top$, one can easily show that $l_{i^{\neq j}}(A^{\neq j})= l_{i_{<j}}(A_{<j}) \cdot l_{i_{>j}}(A_{>j}^\top)$. Sampling from the leverage scores distributions thus boils down to sampling rows of $A_{<j}$ and $A_{>j}^\top$ with probability proportional to
their squared row norms~(due to the orthogonality of $A_{<j}$ and $A_{>j}$ inherited from the canonical form). Without loss of generality, we detail the sampling procedure for $A_{\leq j}$~(the difference between $A_{\leq j}$ and $A_{<j}$ amounts to reindexing). The sampling procedure for $A_{>j}$ will be the same and straightforward.

% Let $\hat s_1, ..., \hat s_j$
% be random variables for indices sampled from $\scr A_1[:,s_1,:],\dots,\scr A_j[:,s_j,:]$.
% We define these variables
% so that the multi-index 
% $\underline{\hat s_1\dots\hat s_j}$ is selected with probability proportional to the squared norm of the
% corresponding row of $A_{\leq j}$. 
% Each variable $\hat s_k$
% takes on values in the set $\br{I_k}$, and since $A_{<j}$ and $A^{{\neq j}^{\top}} A^{\neq j}=I_{R^2}$, the joint probability distribution is obtained by
% \begin{equation}    
% p(\hat s_1 = s_1,\dots,\hat s_j = s_j)
% := \frac{1}{R_j}\paren{A_{\leq j}[\underline{s_1 \dots s_j}, :] \cdot A_{\leq j}[\underline{s_1\dots s_j}, :]^\top}.
% \label{eq:hat_s_defn}
% \end{equation}
Let $\hat s_1\in \br{I_1}, ..., \hat s_j\in \br{I_j}$
be random variables such that the multi-index 
$\hat s_{\leq j} = \underline{\hat s_1\dots\hat s_{j}}$ follows the leverage score distribution of $A_{\leq j}$. 
Since $\TT(({\scr A_n})_{n=1}^N)$ is in canonical form w.r.t. $j+1$, $A_{\leq j}$ is an orthonormal matrix, hence $\underline{\hat s_1\dots\hat s_j}$ is selected with probability proportional to the squared norm of the
corresponding row of $A_{\leq j}$:
\begin{equation}    
p(\hat s_1 = s_1,\dots,\hat s_j = s_j)
:= \frac{1}{R_j}\paren{A_{\leq j}[\underline{s_1 \dots s_j}, :] \cdot A_{\leq j}[\underline{s_1\dots s_j}, :]^\top}.
\label{eq:hat_s_defn}
\end{equation}
Our sampling procedure will draw a lateral slice from each core starting from $\scr A_j$ and ending with $\scr A_1$, corresponding to a single row of $A_{\leq j}$. Suppose we have drawn $s_{k+1}, \dots, s_j,$ for some $k<j$. To sample the $k$-th index, we need to compute the conditional probability $p(s_k\vert s_{k+1},\dots ,s_j)=\frac{p(s_{k},\dots ,s_j)}{p(s_{k+1},\dots ,s_j)}$.
%By starting the sampling procedure at the $j$-th core, we can exploit the left-orthonormality property to derive the conditional distribution of $\hat s_k$ for $k<j$.
The following lemma shows that this can be done efficiently by leveraging the underlying TT structure.
\begin{lemma}[Conditional distribution for $\hat s_k$]
Consider the events
$\hat s_{j} = s_j,\dots, \hat s_{k+1} = s_{k+1}$, which we abbreviate as $\hat s_{>k} = s_{>k}$. Then
\[p(\hat s_k = s_k\ \vert\ \hat s_{>k} = s_{>k})
\propto\mathrm{Tr} \br{H_{>k}^\top \cdot \scr A_k \br{:, s_k, :}^\top \cdot \scr A_k \br{:, s_k, :} \cdot H_{>k}},
\]
where $H_{>k} := \scr A_{k+1} \br{:, s_{k+1}, :} \cdot ... \cdot \scr A_j \br{:, s_j, :}.$
\label{lemma:conditional_distribution}
\end{lemma}
The proof is given in Appendix~\ref{appendix:conditional_lemma_proof}. Intuitively, $H_{>k}$ acts as a ``history matrix" conditioning on $ s_{>k}$, while the trace operation corresponds to marginalization over $s_{<k}$. Unfortunately, updating 
$H_{>k}$ through matrix multiplication as each
index is selected still requires time $O(R^3)$~(assuming $R_1 = ... = R_j = R$). In order to further improve the runtime and reach the quadratic complexity in $R$ claimed in Theorem~\ref{thm:main_result}, we make the following observation: let $q\in\RR^{\prod_{i\leq j }I_i}$ be the
probability vector for the leverage score distribution of $A_{\leq j}$.
Then Equation~\eqref{eq:hat_s_defn} can be rewritten in vector form as
$
q := \frac{1}{R_j} \paren{A_{\leq j}\br{:, 1}^2 + ... + A_{\leq j}\br{:, R_j}^2}.
$
Here, the square of
each column vector is
an elementwise operation. Observe that each $A_{\leq j}\br{:, r}^2$ is a probability vector (positive entries summing to one) due to the orthonormality of $A_{\leq j}$. Hence $q$ is a 
\textit{mixture distribution}. 
To sample from $q$, 
 it thus suffices to select a
single column $\hat r$ of  $A_{\leq j}$ uniformly at random and restrict the
sampling procedure to $A_{\leq j}\br{:, \hat r}^2$. More formally, let $\hat r$ be
uniformly distributed over $\br{R_j}$ and let
$\hat t_1, ..., \hat t_j$ follow the conditional distributions defined by
\begin{equation}
p(\hat t_k = t_k\ \vert\ \hat t_{k+1} = t_{k+1},\dots,\hat t_j = t_j, \hat r = r) = \norm{
\scr A_k \br{:, t_k, :} \cdot h_{>k}}^2,
\label{eq:hat_t_defn}
\end{equation}
where $h_{>k} = \scr A_{k+1}\br{:, t_{k+1}, :} \cdot ... \cdot
\scr A_j \br{:, t_j, r}$. We have 
the following result.
\begin{lemma}\label{lemma:equivalent_conditional_distribution}
For any choice of
$s_j, ..., s_k$, fix $s_j = t_j, s_{j-1} = t_{j-1}, ..., s_k = t_k$.
After marginalizing over $\hat r$, the conditional
distribution of $\hat t_k$ satisfies
$
p(\hat t_k = t_k\ \vert\ \hat t_{>k} = t_{>k}) = 
p(\hat s_k = s_k\ \vert\ \hat s_{>k} = s_{>k}).$
\end{lemma}
As a consequence, the joint random variable 
$(\hat t_1, ..., \hat t_j)$ follows the desired squared
row-norm distribution  of $A_{\leq j}$ after 
marginalizing over $\hat r$. The proof
appears in Appendix \ref{appendix:equivalent_lemma_proof}.
Notice that the ``history matrix" $H_{>k}$
has been replaced by a vector $h_{>k}$. This vector
can be updated by matrix-vector multiplication, yielding
a reduced sampling complexity with only a quadratic dependency on $R$.

Our final improvement is to show that each sample from
the distribution in Equation \eqref{eq:hat_t_defn} can be drawn 
in time sublinear in
the dimension $I_k$ (after appropriate preprocessing). Letting $A^L_k$ be the 
left unfolding of $\scr A_k$, one can check that 
\begin{equation}\label{eq:consecutive_entries}
p(\hat t_k = t_k\ \vert\ \hat t_{>k} = t_{>k},\hat r = r)
%= \norm{A_k^L_{\br{{t_k R_{k-1}}:{(t_k + 1) R_{k-1}}, :}} \cdot h_{>k}}_2^2
= \sum_{i=0}^{R_{k-1}-1} \paren{A_k^L\br{t_k R_{k-1} + i, :} \cdot h_{>k}}^2.
\end{equation}
The probability of selecting the slice $s_k$ is thus the sum of $R_{k-1}$ consecutive entries
from the probability vector $(A^L_k \cdot h_{>k})^2$. As a result, we can sample 
$\hat t_k$ by first sampling an index in the range
$\br{I_k R_{k-1}}$ given by $(A^L_k \cdot h_{>k})^2$, then
performing integer division by $R_{k-1}$ to obtain the corresponding slice index $\hat t_k$. 
The advantage here lies in an efficient data structure for sampling
from the weight vector $(A^L_k \cdot h_{>k})^2$, 
given by the following lemma: 

\begin{lemma}[\citet{bharadwaj2023fast}, Adapted]
Given a matrix $A \in \RR^{I \times R}$, there exists a data structure with construction time
$O(I R^2)$ and space usage $O(IR)$ such that, given any vector $h \in \RR^R$, a single sample from the un-normalized
distribution of weights $(A \cdot h)^2$ can be drawn in time 
$O(R^2 \log (I / R))$.
\label{lemma:efficient_row_sampler}
\end{lemma}
\begin{wrapfigure}{R}{0.5\textwidth}
    \begin{minipage}{0.5\textwidth}
\begin{algorithm}[H]
    \caption{ConstructChainSampler($\scr A_1, ..., \scr A_{N}$)}
    \begin{algorithmic}[1]
    \FOR{$k =1..N$}
        \STATE $Z_k := \textrm{BuildSampler}(A^L_k)$
    \ENDFOR
    \end{algorithmic}
    \label{alg:chain_sampler_construction}
\end{algorithm}
\vspace{-1em}
\begin{algorithm}[H]
    \caption{ChainSampleLeft($J, j$)}
    \begin{algorithmic}[1]
    \FOR{$d =1..J$}
        \STATE $\hat r := \textrm{Uniform-sample}(\br{1...R_j})$
        \STATE $h := e_{\hat r}$ 
        \FOR{$k=j...1$}
            \STATE $\hat t_k := \textrm{RowSample}(Z_k, h) // R_{k-1}$ 
            \STATE $h = h \cdot \scr A_k \br{:, \hat t_k, :}$ 
        \ENDFOR
        \STATE $t_d = (\hat t_k)_{k \leq j}$
    \ENDFOR
    \STATE \textbf{return} $t_1, ..., t_J$
    \end{algorithmic}
    \label{alg:chain_sampling}
\end{algorithm}  
    \end{minipage}
  \end{wrapfigure} 
The adaptation of this lemma is given in Appendix
\ref{appendix:efficient_sampling_ds}. 
Lemma \ref{lemma:efficient_row_sampler} enables us to efficiently draw samples according to the distribution in Equation 
\ref{eq:consecutive_entries}, and therefore 
gives us a procedure 
to sample from the entire core chain. Constructing
the data structure above for each matrix $A^L_k$,
$1 \leq k \leq j$, costs $O(IR_{k-1}R_k^2)$
with a linear space overhead in the input
core sizes. Drawing a sample from the $k$-th
data structure requires time
$O(R_k^2 \log(I_k R_{k-1} / R_{k}))$.~Summing up
this runtime over $1 \leq k \leq j$ gives the 
stated complexity in Theorem \ref{thm:main_result}.
Algorithms 
\ref{alg:chain_sampler_construction} and 
\ref{alg:chain_sampling} summarize the procedures to efficiently
draw $J$ samples from a left-orthogonal core chain. The construction 
procedure builds a set of data structures $Z_k$ of the form given by
Lemma \ref{lemma:efficient_row_sampler} on the left-matricization 
of each tensor core. For each of $J$ rows to draw,
the sampling algorithm selects a column 
$\hat t$ uniformly at random
from the left matricization. It then
initializes the history vector $h$ and successively
samples indices $\hat t_{j-1}, ..., \hat t_1$ 
according to the conditional distribution, 
updating the history vector at each step. 
Appendix \ref{sec:main_proof} provides a 
rigorous proof of the correctness of the procedure
sketched in this section.

While the proof sketched above shares similarities with
the sampler proposed by \cite{bharadwaj2023fast},
key adaptations are required to sample from a tensor 
train core chain. The factors of a Khatri-Rao product
can be sampled in any order, since the Khatri-Rao product
of several matrices is commutative up to a permutation of
its rows. Our sampling procedure \textbf{requires} us
to sample from core $\scr A_j$ down to $\scr A_1$,
since Lemma \ref{lemma:conditional_distribution} exploits
the left-orthogonality of the each core in its derivation.
Starting the sampling procedure at $\scr A_j$ leads
to a ``history matrix" to keep track of prior 
draws instead of the vector that would arise
starting from core $\scr A_1$. Here, our second
innovation of sampling a column uniformly at random
is required to bring down the overall sampling complexity.
We can now state the following guarantee for $\textbf{Randomized-TT-ALS}$~(rTT-ALS) applying the data structure in Theorem~\ref{thm:main_result}. The proof is given in Appendix~\ref{sec:corollary-4.4}.
\begin{corollary}(rTT-ALS)\label{cor:rtt-als-alg}
    %Let $A_{<j}$ and  $A_{>j}$ be the left and right subchain matrices. Let $S\in\RR^{J\times\prod_{k=1}^N I_k}$ be the sampling matrix. 
    For any $\varepsilon,\delta\in (0,1)$ the sampling procedure proposed above guarantees that with $J = \tilde{O}(R^2/\varepsilon\delta)$ samples per least-square problem, we have $$\norm{A^{\neq j}(\tilde{A_j})_{(2)}^\top-X_{(j)}^\top}\leq(1+\varepsilon)\min_{(A_j)_{(2)}}\norm{A^{\neq j}(A_j)_{(2)}^\top- X_{(j)}^\top},$$ with probability $(1-\delta)$, where $\tilde{A_j}$ is the solution of the sketched least-squares problem, for all least-squares solve. The efficient sampling procedure of Theorem~\ref{thm:main_result} brings the overall complexity to
      $\Tilde{O}\paren{\frac{\noiter}{\varepsilon\delta}R^4\cdot\sum_{j=1}^N N\log I_j + I_j},$
where ``$\noiter$" is the number of ALS iterations.
\end{corollary}

\section{Experiments}
% All experiments are run on CPU nodes of the NERSC Perlmutter 
% HPE supercomputer and the Mila Quebec AI Institute compute cluster.
In this section, we demonstrate the effectiveness of the proposed rTT-ALS on two types of tensors:
(i) synthetic and real dense datasets and (ii) real sparse datasets. We use the fit as evaluation metric~(higher is better):  $\text{fit}(\Tilde{\scr X},\scr X) = 1-{\|\Tilde{\scr X}-\scr X\|_F}/{\|\scr X\|_F}$,  where $\tilde{\scr X}$ is the TT approximation and $\scr X$ is the target tensor.
\subsection{Decomposition of Synthetic and Real Dense Datasets}
We compare {rTT-ALS} to three other methods; TT-SVD~\citep{oseledets2011tensor}, Randomized TT-SVD (rTT-SVD)~\citep{huber2017randomized} and TT-ALS~\citep{holtz2012alternating}. We use TensorLy~\citep{tensorly} for SVD-based methods and our own implementation for deterministic TT-ALS. For simplicity, we set $R_1=\dots=R_{N-1} = R$ for all experiments. 
%We also choose all decomposition ranks equal for each synthetic and real datasets experiments. 
For all algorithms, we illustrate the quality of performance by fit and runtime. 

\begin{figure*}
    \centering
    \includegraphics[width=0.3\textwidth]{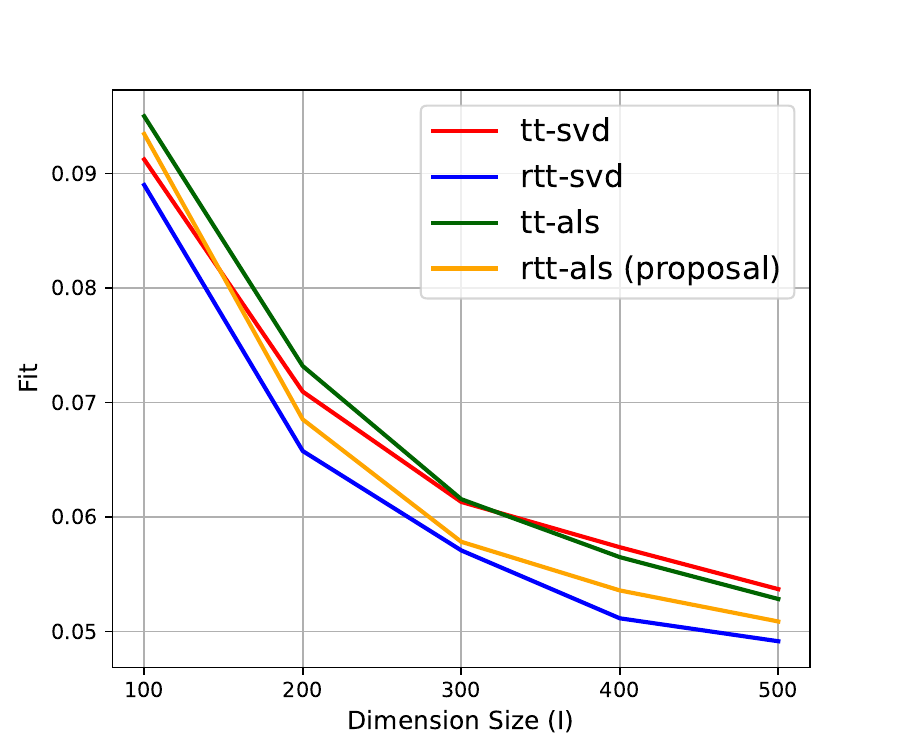} \hspace{3cm}\includegraphics[width=0.3\textwidth]{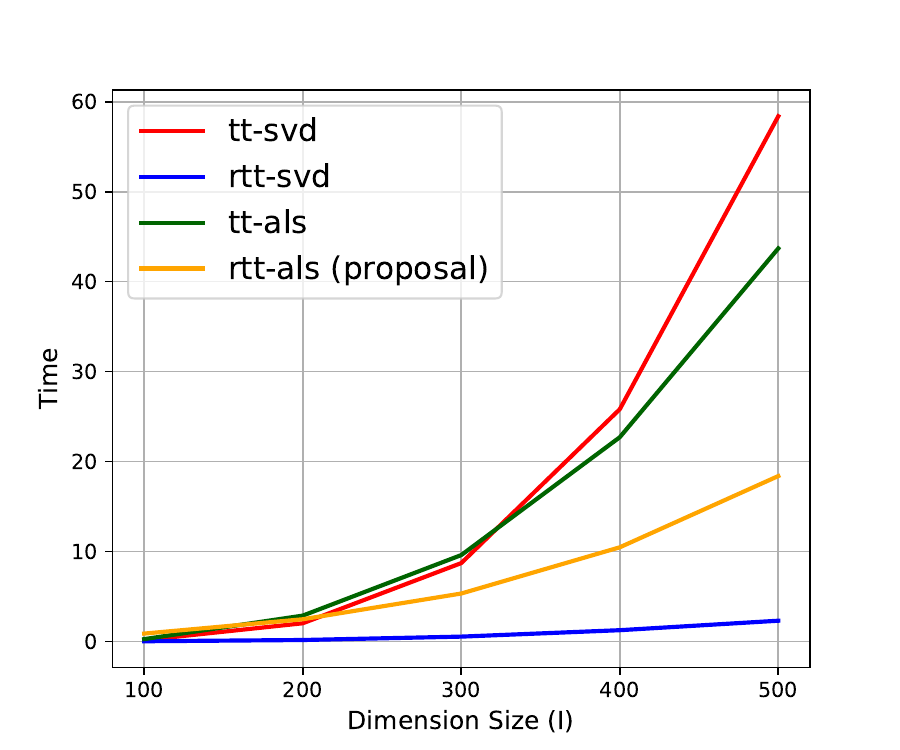}
    \caption{
    Fit (left) and running time (right) averaged over 5 trials for the synthetic data experiment.
    %Synthetic data experiments for four methods for $R=20$ and $\Tilde{R}=5$. The sample count is set to $J=5000$. (left) Fit and (right) time for the average of 5 trials are reported.
    }\label{fig:synthetic}
\end{figure*}

\textbf{Synthetic Data Experiments.}~
For the synthetic data experiment, we generate random tensors of size $I \times I \times I$ for $I\in\{100,\dots,500\}$ and of TT rank $R=20$~(by drawing each core's components i.i.d. from a standard normal distribution). A small Gaussian noise with mean zero and standard deviation of $10^{-6}$ is added to each entry of the resulting tensor. We then run the four methods to find a rank $\tilde{R}=5$ approximation of the target tensor. ALS-based methods are initialized using their SVD-based counterpart~(TT-ALS with the output of TT-SVD and rTT-ALS with the output of rTT-SVD) and are run for 15 iterations. The sample count for rTT-ALS is fixed to $J =5000$ for all values of $I$. The average fit over 5 trials for all four algorithms are reported as a function of the dimension in Figure~\ref{fig:synthetic}. 
rTT-ALS is about $2\times$ faster than TT-ALS and $3\times$ faster than TT-SVD for $I=500$. Although rTT-SVD  is the fastest method, it achieves poor performance in terms of fit.

\textbf{Real Data Experiments.}~
For the real data experiment, we consider four real images and video datasets~(more details about datasets are given in Appendix~\ref{app:datasets}): (i) Pavia University is a hyper-spectral image dataset of size $(610 \times 340 \times 103)$, (ii) DC Mall is also a dataset of hyper-spectral images of size $(1280 \times 307 \times 191)$. Both datasets are three-dimensional tensors where the first two dimensions are the image height and width, and the third dimension is the number of spectral bands, (iii) the MNIST dataset is of size $(60000\times 28\times 28)$, and iv) Tabby Cat is the three-dimensional tensor of size $(720 \times 1280 \times 286)$ which contains grayscale videos of a man sitting on a park bench and a cat, respectively. The first two dimensions are frame height and width, and the third dimension is the number of frames. For all datasets, the preprocessing step is done by tensorizing data tensors into higher-dimensional tensors. Table~\ref{tab:fit-real-data} illustrates the results for a single trial when $\Tilde{R}=5$. For all datasets we keep the sample count fixed to $J=2000$. Similarly to the synthetic data experiments, rTT-ALS is faster than TT-ALS and TT-SVD~(up to $10\times$ faster than TT-ALS). 
% Even though rTT-SVD yields faster speedup compared to rTT-ALS, it achieves a slightly worse fit.

% \begin{table}[h]
% \centering
% \caption{Runtime, decomposition error and accuracy when using tensor decomposition}
% \label{tab:results}
% \begin{tabular}{@{}lccc@{}}
% \toprule
% Method                  & Time (s) & Error & Acc\\ \midrule
% TT-SVD   & 3.56  & 0.388 & 0.919   \\
% rTT-SVD    & 1.43  & 0.388 & 0.918    \\
% TT-ALS &26.4 & 0.387 & 92.5\\
% rTT-ALS~(proposal)   & 1.66& 0.389 & 0.916
% \\ \bottomrule
% \end{tabular}
% \end{table}
\sisetup{round-mode=places, table-number-alignment=right, table-text-alignment=right}

\begin{table*}
\vspace*{-0.2cm}
	\centering
	\caption{Decomposition results for real datasets with target rank $\Tilde{R}=5$.
	Time is in seconds.} 
	\begin{tabular}{
			l
			S[round-precision=2, table-figures-decimal=2, table-figures-integer=1]
			S[round-precision=2, table-figures-decimal=12, table-figures-integer=2]
			S[round-precision=2, table-figures-decimal=2, table-figures-integer=1]
			S[round-precision=3, table-figures-decimal=3, table-figures-integer=3]
			S[round-precision=2, table-figures-decimal=2, table-figures-integer=1]
			S[round-precision=2, table-figures-decimal=2, table-figures-integer=4]
			S[round-precision=2, table-figures-decimal=2, table-figures-integer=1]
			S[round-precision=2, table-figures-decimal=2, table-figures-integer=4]
			S[round-precision=2, table-figures-decimal=2, table-figures-integer=1]
			S[round-precision=2, table-figures-decimal=2, table-figures-integer=3]
		}  
		\toprule
		& \multicolumn{2}{c}{Pavia Uni.}
            &\multicolumn{2}{c}{Tabby Cat}
            &\multicolumn{2}{c}{MNIST}
             &\multicolumn{2}{c}{DC Mall}\\
		\cmidrule(lr){2-3}
		\cmidrule(lr){4-5}
            \cmidrule(lr){6-7}
            \cmidrule(lr){8-9}
            \cmidrule(lr){10-11}
		Method 				 & {Fit} & {Time} & {Fit} & {Time} & {Fit} & {Time}  & {Fit} & {Time}\\
		\midrule
		TT-ALS 				 
         & 0.6103819274363476
         &4.159196376800537  
         & 0.653903489079781 
         &44.56976294517517
         &0.4581272148325476
         &8.28534340858459
         &0.5928250170733753 
         &21.856302976608276\\
	\textbf{rTT-ALS~(proposal)} 			 
        &0.6045745873636315
        &0.8219459056854248
         &0.6492514805833833 
         &7.35988712310791 
         &0.45045158424742504 
         &2.196021318435669
         &0.5872824342039946 
         &2.8072521686553955\\
	\midrule
		TT-SVD 				 
        &0.610367476940155
        &6.6491944789886475
        &0.6538996044843575  
        &136.18942737579346
        & 0.4581217355264319
        &17.19279718399048
        & 0.5919069945812225
        &41.44852805137634\\
		rTT-SVD		 
        &0.608104944229126
        &0.3314323425292969
        & 0.6499373963266095   
        &4.284693241119385
        &0.45807390472586873  
        &0.652860164642334
        &0.5919323265552521  &0.46129822731018066\\
		\bottomrule
	\end{tabular}\label{tab:fit-real-data}
\end{table*}
\subsection{Approximate Sparse Tensor 
Train Decomposition}
\begin{figure}
     \centering
     \begin{subfigure}[b]{0.32\textwidth}
         \centering
         \includegraphics[scale=0.37]{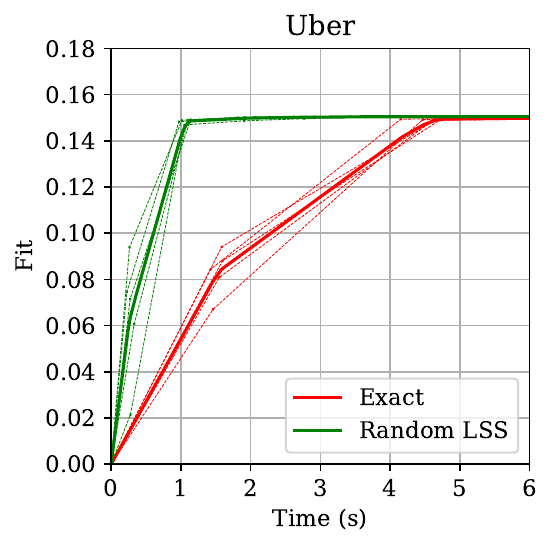}
     \end{subfigure}
     \hfill
     \begin{subfigure}[b]{0.32\textwidth}
         \centering
         \includegraphics[scale=0.37]{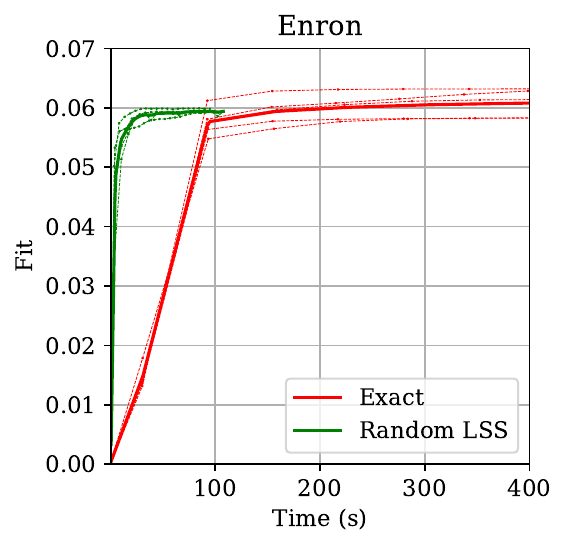}
     \end{subfigure}
     \begin{subfigure}[b]{0.32\textwidth}
         \centering
         \includegraphics[scale=0.37]{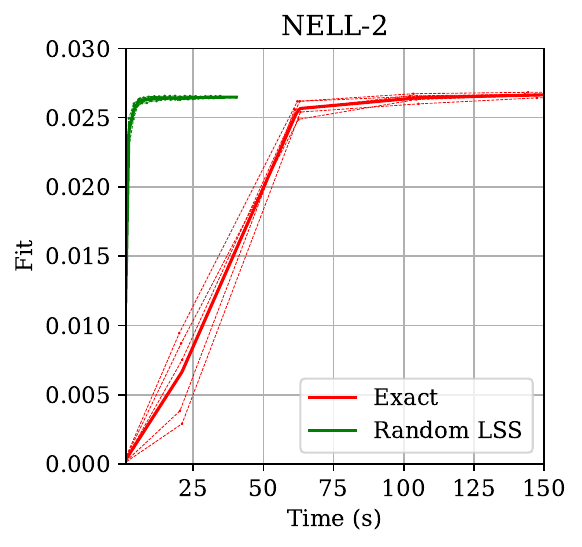}
     \end{subfigure} 
    \caption{Fit as a function of time for three FROSTT
    tensors, $R=6$, $J=2^{16}$ for rTT-ALS. Thick lines are averages of 5 fit-time 
    traces, shown by thin dotted lines.}
    \label{fig:fit_function_time}
\end{figure}

We next apply  rTT-ALS  to three large sparse tensors from FROSTT \citep{smith_frostt_2017}. Table \ref{tab:sparse_tensor_fits} gives the
fits achieved by our 
method to decompose these tensors. The largest of these tensors,
NELL-2, has around 77 million nonzero entries with mode sizes
in the tens of thousands. Fits for sparse tensor decomposition
are typically low, but the factors of the resulting 
decomposition have successfully been mined for patterns
\citep{larsen2022practical}. For these experiments,
we chose all decomposition ranks equal with
$R_1 = ... = R_N = R$ and tested over a
range of values for $R$.
\begin{table*}
\centering
\caption{Average fits and speedup, $J=2^{16}$ for
randomized algorithms, 40 ALS
iterations. The speedup  is the average per-iteration runtime
for a single exact ALS sweep divided by the average time
for a single randomized sweep.}
% \scalebox{0.9}{
\begin{tabular}{l|ccc|ccc|ccc}
\toprule
\multicolumn{1}{c}{}& \multicolumn{3}{c}{Uber}& \multicolumn{3}{c}{Enron}& \multicolumn{3}{c}{NELL-2}\\
\midrule
$R$  & rTT-ALS  & TT-ALS  &  Speedup & rTT-ALS  & TT-ALS  &  Speedup & rTT-ALS  & TT-ALS  &  Speedup\\
% of LSS over Exact \\
\midrule
4   & 0.1332 & 0.1334 & 4.0x & 0.0498 & 0.0507 & 17.8x & 0.0213 & 0.0214 & 26.0x\\ 
6   & 0.1505 & 0.1510 & 3.5x & 0.0594 & 0.0611 & 12.4x& 0.0265 & 0.0269 & 22.8x \\ 
8   & 0.1646 & 0.1654 & 3.0x & 0.0669 & 0.0711 & 10.5x & 0.0311 & 0.0317 & 22.2x\\ 
10   & 0.1747 & 0.1760 & 2.4x & 0.0728 & 0.0771 & 8.5x  & 0.0350 & 0.0359 & 20.5x \\ 
12   & 0.1828 & 0.1846 & 1.5x  & 0.0810 & 0.0856 & 7.4x  & 0.0382 & 0.0394 & 15.8x \\
\bottomrule
% \multicolumn{1}{c}{}
% \multirow{5}{*}{Uber}   
% & 4   & 0.1332 & 0.1334 & 4.0x \\ 
% & 6   & 0.1505 & 0.1510 & 3.5x \\ 
% & 8   & 0.1646 & 0.1654 & 3.0x \\ 
% & 10   & 0.1747 & 0.1760 & 2.4x \\ 
% & 12   & 0.1828 & 0.1846 & 1.5x \\ 
% \midrule
% \multirow{5}{*}{Enron}   
% & 4   & 0.0498 & 0.0507 & 17.8x \\ 
% & 6   & 0.0594 & 0.0611 & 12.4x \\ 
% & 8   & 0.0669 & 0.0711 & 10.5x \\ 
% & 10   & 0.0728 & 0.0771 & 8.5x \\ 
% & 12   & 0.0810 & 0.0856 & 7.4x \\ 
% \midrule
% \multirow{5}{*}{NELL-2}   
% & 4   & 0.0213 & 0.0214 & 26.0x \\ 
% & 6   & 0.0265 & 0.0269 & 22.8x \\ 
% & 8   & 0.0311 & 0.0317 & 22.2x \\ 
% & 10   & 0.0350 & 0.0359 & 20.5x \\ 
% & 12   & 0.0382 & 0.0394 & 15.8x \\ 
% \bottomrule
\end{tabular}\label{tab:sparse_tensor_fits}
\end{table*}

The fits produced by rTT-ALS match those produced by the 
non-randomized ALS method up to variation in the third
significant figure for Uber and NELL-2, with slightly higher
errors on the Enron tensor. We kept the sample count for
our randomized algorithms fixed at $J=2^{16}$ throughout 
this experiment. As a result, the gap between the fit
of the randomized and exact methods grows as the target rank 
increases, which our theory predicts. 

Table~\ref{tab:sparse_tensor_fits} also reports
the average speedup per ALS sweep of rTT-ALS over the exact algorithm. On the NELL-2 sparse tensor with target rank 12, the non-randomized ALS algorithm requires an average of 29.4 seconds per
ALS sweep, while rTT-ALS requires only 
1.87 seconds. Figure \ref{fig:fit_function_time} 
shows that our method makes faster progress than its 
non-randomized counterpart across all three tensors. Because
we could not find a well-documented, high-performance library
for sparse tensor train decomposition, we wrote a fast 
multithreaded implementation in C++, which serves as the baseline
method in these figures and tables~(the code will be made publicly available).
%
%
%
% \begin{table}[h]
% \centering
% \begin{tabular}{llccc}
% \toprule
% Tensor & $R$ & rTT-ALS Fit & Exact ALS Fit & Avg. Speedup\\
% % of LSS over Exact \\
% \midrule
% \input{Styles/tables/sparse_tensor_fits}
% \end{tabular}
% \caption{Average Fits and speedup, $J=2^{16}$ for
% randomized algorithms, 40 ALS
% iterations. The speedup of leverage-score
% sampling is the average per-iteration runtime
% for a single exact ALS sweep divided by the average time
% for a single randomized sweep.}
% \label{tab:sparse_tensor_fits}
% \end{table}

\medskip

\begin{figure}
    \centering
    \vspace*{0.4cm}
    \includegraphics[scale=0.3]{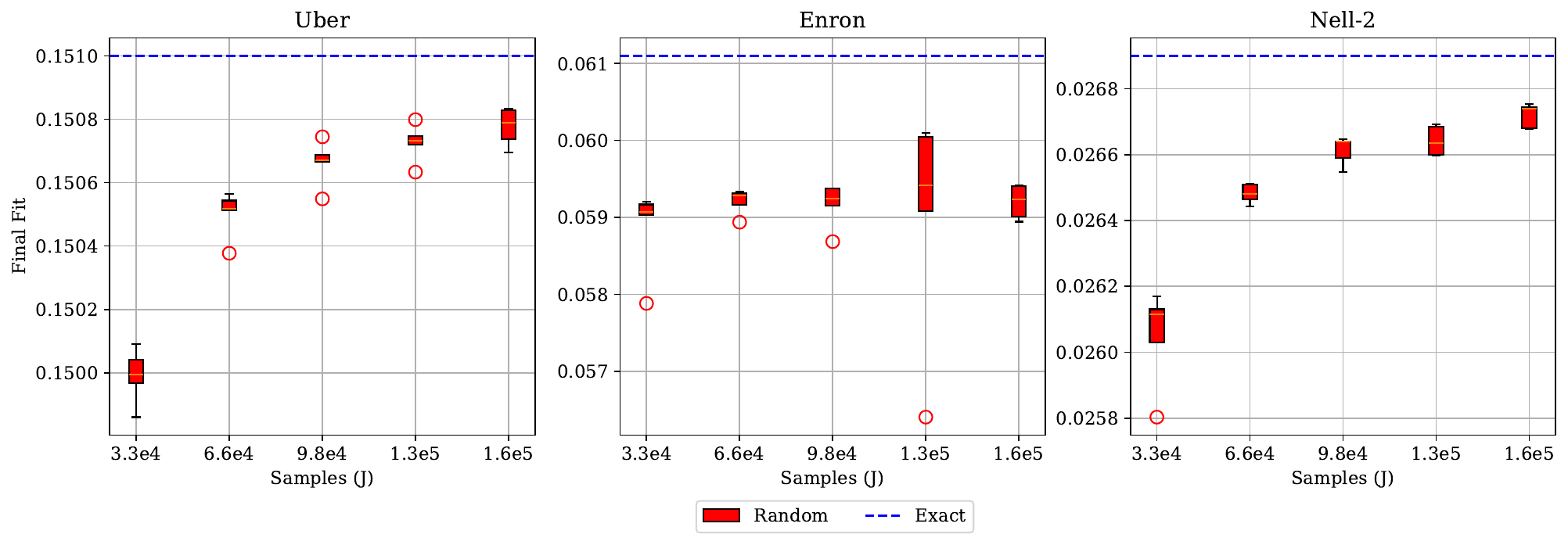}
    \caption{Final fit of sparse tensor decomposition for varying
    sample counts. Each boxplot reports statistics for
    5 trials. The blue dashed lines show the fit for
    non-randomized ALS.}
    \label{fig:accuracy_vs_sample_count}
\end{figure}
Figure \ref{fig:accuracy_vs_sample_count} shows the impact
of varying the sample count on the final fit. We find modest increases in accuracy for both
Uber and NELL-2 as the sample count increases by a factor of 5
(starting from $J=2^{15}$). Increasing $J$
has a smaller impact for the Enron tensor, which is generally
more difficult to decompose beginning with i.i.d. random
factor initialization~\citep{larsen2022practical}.

\section{Conclusion}
We proposed a sampling-based ALS method leveraging an efficient data structure to sample from the exact leverage scores. More precisely, we show that by exploiting the canonical form of the TT decomposition, leverage scores can be computed efficiently for all the least squares problems of ALS. We provide strong theoretical guarantees for the proposed data structure. Experiments on massive dense and sparse tensors confirm the theoretical results.
The sampling algorithm we proposed could be extended to more general tree-based tensor network structures, leveraging canonical forms in a similar spirit to rTT-ALS. 
\bibliography{biblio}
\newpage
\clearpage
\newpage
\appendix

\section{Additional Notations}\label{appendix:additinal-notations}
\begin{definition}\label{kron}
Let $A\in\RR^{m\times n}$ and $B\in\RR^{p\times q}$ then the Kronecker product, $A\otimes B\in\RR^{mp\times nq}$ is defined by
\begin{align*}
    \def\baseline{-0.5ex}
    A\otimes B = 
    \begin{bmatrix}
    a_{11}B & a_{12}B & \dots & a_{1n}B \\
    a_{21}B & a_{22}B & \dots & a_{2n}B \\
    \vdots & \vdots & \ddots & \vdots \\
    a_{m1}B & a_{m2}B & \dots & a_{mn}B
    \end{bmatrix}
\end{align*}
\end{definition}
\begin{definition}
Let $A\in\RR^{m\times R}$ and $B\in\RR^{n\times R}$ then the Khatri-Rao product, $A\odot B\in\RR^{mn\times R}$ is defined by
\begin{align*}
    \def\baseline{-0.5ex}
    A\odot B =
    \begin{bmatrix}
    \vline & \vline  &   & \vline \\
    a_1\otimes b_1 & a_2\otimes b_2 & \dots & a_R\otimes b_R\\
    \vline & \vline & & \vline 
    \end{bmatrix}
\end{align*}
where $a_1,\dots,a_R\in\RR^{m}$ are the columns of $A$, $b_1,\dots,b_R\in\RR^{n}$ are the columns of $B$ and the columns of $A\odot B$ is the subset of the Kronecker product. In the corresponding tensor network diagram, the copy tensor captures the fact that the second indices are the same. 
\end{definition}
\subsection{Details about Orthogonalization of the TT Decomposition}\label{app:orthogonal-tt}
\begin{figure}[h!]
    \centering
\begin{tikzpicture}[baseline=-0.5ex]
    \tikzset{tensor/.style = {minimum size = 0.5cm,shape = circle,thick,draw=black,inner sep = 0pt}, edge/.style = {thick,line width=.4mm},every loop/.style={}}
    \def\x{0}
    \def\y{0}
    \node[tensor, 
        fill=red!30!white, shift = {(\x-4,\y)}] (A) {\scalebox{0.85}{$\scr A_1$}};
        \draw[edge](A) -- (\x-4,\y-0.6) node [midway,left] {\scalebox{0.8}{\textcolor{gray}{$i_1$}}};
       \node[tensor, path picture={
        \fill[fill=red!50!white](path picture bounding box.south west)
        -- 
        (path picture bounding box.north east) --
        (path picture bounding box.north west) -- cycle;
        }, shift={(\x-3,\y)}] (B) {\scalebox{0.85}{$\scr A_2$}};
       \draw[edge](B) -- (\x-3,\y-0.6)node [midway,left] {\scalebox{0.7}{\textcolor{gray}{$i_2$}}};
        \node[tensor, path picture={
        \fill[fill=blue!50!white](path picture bounding box.south west)
        -- 
        (path picture bounding box.north east) --
        (path picture bounding box.north west) -- cycle;
        }, shift={(\x-2,\y)}] (C) {\scalebox{0.85}{$\scr A_3$}};
        \draw[edge](C) -- (\x-2,\y-0.6)node [midway,left] {\scalebox{0.7}{\textcolor{gray}{$i_3$}}};
        \node[tensor, path picture={
       \fill[fill=blue!60!red!50!](path             picture bounding box.south west)
        -- 
        (path picture bounding box.north east) --
       (path picture bounding box.north west) -- cycle;}, shift={(\x-1,\y)}] (D) {\scalebox{0.85}{$\scr A_4$}};
    \draw[edge](D) -- (\x-1,\y-0.6)node [midway,left] {\scalebox{0.7}{\textcolor{gray}{$i_4$}}};
    \node[tensor, path picture={
    \fill[fill= blue!20!white](path             picture bounding box.south west)
    -- 
    (path picture bounding box.north east) --
    (path picture bounding box.north west) -- cycle;}, shift={(\x,\y)}] (E) {\scalebox{0.85}{$\scr A_5$}};
    \draw[edge](E) -- (\x,\y-0.6)node [midway,left] {\scalebox{0.7}{\textcolor{gray}{$i_5$}}};
    \draw[edge](A) -- (B) node [midway,above] {\scalebox{0.7}{\textcolor{gray}{$R_1$}}};
    \draw[edge](B) -- (C) node [midway,above] {\scalebox{0.7}{\textcolor{gray}{$R_2$}}};
    \draw[edge](C) -- (D) node [midway,above] {\scalebox{0.7}{\textcolor{gray}{$R_3$}}};
    \draw[edge](D) -- (E) node [midway,above] {\scalebox{0.7}{\textcolor{gray}{$R_4$}}};
\end{tikzpicture}~~~\text{step:~1}\\
\begin{tikzpicture}[baseline=-0.5ex]
    \tikzset{tensor/.style = {minimum size = 0.5cm,shape = circle,thick,draw=black,inner sep = 0pt}, edge/.style = {thick,line width=.4mm},every loop/.style={}}
    \def\x{0}
    \def\y{0}
    \node[tensor, path picture = {
        \fill[fill=red!30!white] (path picture bounding box.south east)
        -- 
        (path picture bounding box.north west) --
        (path picture bounding box.north east) -- cycle;
        }, shift = {(\x-4,\y)}] (A) {\scalebox{0.85}{$\scr A_1$}};
        \draw[edge](A) -- (\x-4,\y-0.6) node [midway,left] {\scalebox{0.8}{\textcolor{gray}{$i_1$}}};
         \draw  node[fill = black,circle,inner sep=0pt,minimum size=6pt] (m) at (\x-3.5,\y) {};
       \node[tensor, path picture={
        \fill[fill=red!50!white](path picture bounding box.south west)
        -- 
        (path picture bounding box.north east) --
        (path picture bounding box.north west) -- cycle;
        }, shift={(\x-3,\y)}] (B) {\scalebox{0.85}{$\scr A_2$}};
       \draw[edge](B) -- (\x-3,\y-0.6)node [midway,left] {\scalebox{0.7}{\textcolor{gray}{$i_2$}}};
        \node[tensor, path picture={
        \fill[fill=blue!50!white](path picture bounding box.south west)
        -- 
        (path picture bounding box.north east) --
        (path picture bounding box.north west) -- cycle;
        }, shift={(\x-2,\y)}] (C) {\scalebox{0.85}{$\scr A_3$}};
        \draw[edge](C) -- (\x-2,\y-0.6)node [midway,left] {\scalebox{0.7}{\textcolor{gray}{$i_3$}}};
        \node[tensor, path picture={
       \fill[fill=blue!60!red!50!](path             picture bounding box.south west)
        -- 
        (path picture bounding box.north east) --
       (path picture bounding box.north west) -- cycle;}, shift={(\x-1,\y)}] (D) {\scalebox{0.85}{$\scr A_4$}};
    \draw[edge](D) -- (\x-1,\y-0.6)node [midway,left] {\scalebox{0.7}{\textcolor{gray}{$i_4$}}};
    \node[tensor, path picture={
    \fill[fill= blue!20!white](path             picture bounding box.south west)
    -- 
    (path picture bounding box.north east) --
    (path picture bounding box.north west) -- cycle;}, shift={(\x,\y)}] (E) {\scalebox{0.85}{$\scr A_5$}};
    \draw[edge](E) -- (\x,\y-0.6)node [midway,left] {\scalebox{0.7}{\textcolor{gray}{$i_5$}}};
    \draw[edge](A) -- (B) node [midway,above] {\scalebox{0.7}{\textcolor{gray}{$R_1$}}};
    \draw[edge](B) -- (C) node [midway,above] {\scalebox{0.7}{\textcolor{gray}{$R_2$}}};
    \draw[edge](C) -- (D) node [midway,above] {\scalebox{0.7}{\textcolor{gray}{$R_3$}}};
    \draw[edge](D) -- (E) node [midway,above] {\scalebox{0.7}{\textcolor{gray}{$R_4$}}};
\end{tikzpicture}~~~~~~\text{QR}\\
\begin{tikzpicture}[baseline=-0.5ex]
    \tikzset{tensor/.style = {minimum size = 0.5cm,shape = circle,thick,draw=black,inner sep = 0pt}, edge/.style = {thick,line width=.4mm},every loop/.style={}}
    \def\x{0}
    \def\y{0}
    \node[tensor, path picture = {
        \fill[fill=red!30!white] (path picture bounding box.south east)
        -- 
        (path picture bounding box.north west) --
        (path picture bounding box.north east) -- cycle;
        }, shift = {(\x-4,\y)}] (A) {\scalebox{0.85}{$\scr A_1$}};
        \draw[edge](A) -- (\x-4,\y-0.6) node [midway,left] {\scalebox{0.8}{\textcolor{gray}{$i_1$}}};
       \node[tensor, 
        fill=red!50!white, shift = {(\x-3,\y)}] (B) {\scalebox{0.85}{$\scr A_2$}};
       \draw[edge](B) -- (\x-3,\y-0.6)node [midway,left] {\scalebox{0.7}{\textcolor{gray}{$i_2$}}};
        \node[tensor, path picture={
        \fill[fill=blue!50!white](path picture bounding box.south west)
        -- 
        (path picture bounding box.north east) --
        (path picture bounding box.north west) -- cycle;
        }, shift={(\x-2,\y)}] (C) {\scalebox{0.85}{$\scr A_3$}};
        \draw[edge](C) -- (\x-2,\y-0.6)node [midway,left] {\scalebox{0.7}{\textcolor{gray}{$i_3$}}};
        \node[tensor, path picture={
       \fill[fill=blue!60!red!50!](path             picture bounding box.south west) -- (path picture bounding box.north east) -- (path picture bounding box.north west) -- cycle;}, shift={(\x-1,\y)}] (D) {\scalebox{0.85}{$\scr A_4$}};
    \draw[edge](D) -- (\x-1,\y-0.6)node [midway,left] {\scalebox{0.7}{\textcolor{gray}{$i_4$}}};
    \node[tensor, path picture={
    \fill[fill= blue!20!white](path picture bounding box.south west)
    -- 
    (path picture bounding box.north east) --
    (path picture bounding box.north west) -- cycle;}, shift={(\x,\y)}] (E) {\scalebox{0.85}{$\scr A_5$}};
    \draw[edge](E) -- (\x,\y-0.6)node [midway,left] {\scalebox{0.7}{\textcolor{gray}{$i_5$}}};
    \draw[edge](A) -- (B) node [midway,above] {\scalebox{0.7}{\textcolor{gray}{$R_1$}}};
    \draw[edge](B) -- (C) node [midway,above] {\scalebox{0.7}{\textcolor{gray}{$R_2$}}};
    \draw[edge](C) -- (D) node [midway,above] {\scalebox{0.7}{\textcolor{gray}{$R_3$}}};
    \draw[edge](D) -- (E) node [midway,above] {\scalebox{0.7}{\textcolor{gray}{$R_4$}}};
\end{tikzpicture}~~~\text{step:~2}\\
\begin{tikzpicture}[baseline=-0.5ex]
    \tikzset{tensor/.style = {minimum size = 0.5cm,shape = circle,thick,draw=black,inner sep = 0pt}, edge/.style = {thick,line width=.4mm},every loop/.style={}}
    \def\x{0}
    \def\y{0}
    \node[tensor, path picture = {
        \fill[fill=red!30!white] (path picture bounding box.south east)
        -- 
        (path picture bounding box.north west) --
        (path picture bounding box.north east) -- cycle;
        }, shift = {(\x-4,\y)}] (A) {\scalebox{0.85}{$\scr A_1$}};
        \draw[edge](A) -- (\x-4,\y-0.6) node [midway,left] {\scalebox{0.8}{\textcolor{gray}{$i_1$}}};
       \node[tensor, path picture={
        \fill[fill=red!50!white](path picture bounding box.south east)
        -- 
        (path picture bounding box.north west) --
        (path picture bounding box.north east) -- cycle;
        }, shift={(\x-3,\y)}] (B) {\scalebox{0.85}{$\scr A_2$}};
       \draw[edge](B) -- (\x-3,\y-0.6)node [midway,left] {\scalebox{0.7}{\textcolor{gray}{$i_2$}}};
        \draw  node[fill = black,circle,inner sep=0pt,minimum size=6pt] (m) at (\x-2.5,\y) {};
        \node[tensor, path picture={
        \fill[fill=blue!50!white](path picture bounding box.south west)
        -- 
        (path picture bounding box.north east) --
        (path picture bounding box.north west) -- cycle;
        }, shift={(\x-2,\y)}] (C) {\scalebox{0.85}{$\scr A_3$}};
        \draw[edge](C) -- (\x-2,\y-0.6)node [midway,left] {\scalebox{0.7}{\textcolor{gray}{$i_3$}}};
        \node[tensor, path picture={
       \fill[fill=blue!60!red!50!](path             picture bounding box.south west)
        -- 
        (path picture bounding box.north east) --
       (path picture bounding box.north west) -- cycle;}, shift={(\x-1,\y)}] (D) {\scalebox{0.85}{$\scr A_4$}};
    \draw[edge](D) -- (\x-1,\y-0.6)node [midway,left] {\scalebox{0.7}{\textcolor{gray}{$i_4$}}};
    \node[tensor, path picture={
    \fill[fill= blue!20!white](path             picture bounding box.south west)
    -- 
    (path picture bounding box.north east) --
    (path picture bounding box.north west) -- cycle;}, shift={(\x,\y)}] (E) {\scalebox{0.85}{$\scr A_5$}};
    \draw[edge](E) -- (\x,\y-0.6)node [midway,left] {\scalebox{0.7}{\textcolor{gray}{$i_5$}}};
    \draw[edge](A) -- (B) node [midway,above] {\scalebox{0.7}{\textcolor{gray}{$R_1$}}};
    \draw[edge](B) -- (C) node [midway,above] {\scalebox{0.7}{\textcolor{gray}{$R_2$}}};
    \draw[edge](C) -- (D) node [midway,above] {\scalebox{0.7}{\textcolor{gray}{$R_3$}}};
    \draw[edge](D) -- (E) node [midway,above] {\scalebox{0.7}{\textcolor{gray}{$R_4$}}};
\end{tikzpicture}~~~~~~\text{QR}\\
\begin{tikzpicture}[baseline=-0.5ex]
    \tikzset{tensor/.style = {minimum size = 0.5cm,shape = circle,thick,draw=black,inner sep = 0pt}, edge/.style = {thick,line width=.4mm},every loop/.style={}}
    \def\x{0}
    \def\y{0}
 \node[tensor, path picture={
        \fill[fill=red!30!white](path picture bounding box.south east)
        -- 
        (path picture bounding box.north west) --
        (path picture bounding box.north east) -- cycle;
        }, shift={(\x,\y)}] (A) {\scalebox{0.85}{$\scr A_1$}};
        \draw[edge](A) -- (\x,\y-0.7) node [midway,left] {\scalebox{0.7}{\textcolor{gray}{$i_1$}}};
       \node[tensor, path picture={
        \fill[fill=red!50!white](path picture bounding box.south east)
        -- 
        (path picture bounding box.north west) --
        (path picture bounding box.north east) -- cycle;
        }, shift={(\x+1,\y)}] (B) {\scalebox{0.85}{$\scr A_2$}};
         \draw[edge](B) -- (\x+1,\y-0.7)node [midway,left] {\scalebox{0.7}{\textcolor{gray}{$i_2$}}};
         \node[tensor,fill=blue!50!white] (C) at (\x+2,\y){$\scr A_3$};
          \draw[edge](C) -- (\x+2,\y-0.7)node [midway,left] {\scalebox{0.7}{\textcolor{gray}{$i_3$}}};
          \node[tensor, path picture={
        \fill[fill=blue!60!red!50!](path picture bounding box.south west)
        -- 
        (path picture bounding box.north east) --
        (path picture bounding box.north west) -- cycle;
        }, shift={(\x+3,\y)}] (D) {\scalebox{0.85}{$\scr A_4$}};
         \draw[edge](D) -- (\x+3,\y-0.7)node [midway,left] {\scalebox{0.7}{\textcolor{gray}{$i_4$}}};
           \node[tensor, path picture={
        \fill[fill=blue!20!white](path picture bounding box.south west)
        -- 
        (path picture bounding box.north east) --
        (path picture bounding box.north west) -- cycle;
        }, shift={(\x+4,\y)}] (E) {\scalebox{0.85}{$\scr A_5$}};
         \draw[edge](E) -- (\x+4,\y-0.7)node [midway,left] {\scalebox{0.7}{\textcolor{gray}{$i_5$}}};
    \draw[edge](A) -- (B) node [midway,above] {\scalebox{0.7}{\textcolor{gray}{$R_1$}}};
    \draw[edge](B) -- (C) node [midway,above] {\scalebox{0.7}{\textcolor{gray}{$R_2$}}};
    \draw[edge](C) -- (D) node [midway,above] {\scalebox{0.7}{\textcolor{gray}{$R_3$}}};
    \draw[edge](D) -- (E) node [midway,above] {\scalebox{0.7}{\textcolor{gray}{$R_4$}}};
\end{tikzpicture}~~~\text{step:~3}
\caption{Half-sweep of TT-ALS. In each non-QR step the fully colored core is optimized and in each QR step the non-orthogonal component (depicted by black circle)~is absorbed to the next core. This procedure repeats until reaching the right side of the decomposition then the same procedure is repeated from right until reaching to the left side (not demonstrated in this figure.)}\label{fig:ortho-tt-als-alg}
\end{figure}
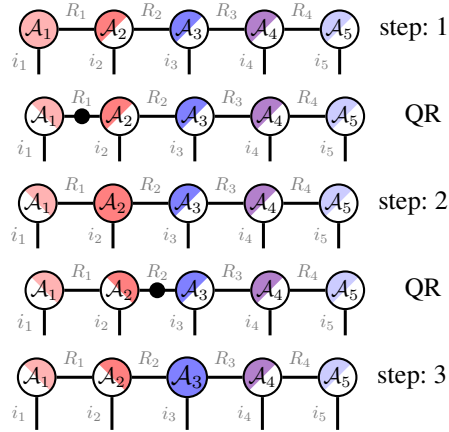

Figure~\ref{fig:ortho-tt-als-alg} illustrates the single-site TT-ALS method, which begins with a TT decomposition in canonical form initialized by a crude guess. 
Core $\scr A_1$ of the decomposition is non-orthogonal; in sweeps from left-to-right and right-to-left,
the algorithm holds all but one core constant and solves for the optimal value for the
remaining core. After updating each core, by a QR decomposition the non-orthonormal part is merged to the 
left or right~(depending on the direction of the sweep),
a step which is called \textit{core orthogonalization}.

\section{Proofs}\label{appendix:conditional_lemma_proof}
\subsection{Proof of Lemma \ref{lemma:conditional_distribution}}
\begin{proof}[\unskip\nopunct]
Noting that $A_{\leq j}[\underline{s_1 \dots s_j}, :]$ is
a \textbf{row vector}, we write
\begin{equation}
\begin{aligned}
p(&\hat s_k = s_k\ \vert\ \hat s_{>k} = s_{>k}) \\
&= 
\sum_{s_1, ..., s_{k-1}}
p(\hat s_1 = s_1 \wedge ... \wedge \hat s_j = s_j) \\
&= \sum_{s_1, ..., s_{k-1}}
\frac{1}{R_j} \left(A_{\leq j}[\underline{s_1 \dots s_j}, :] \cdot A_{\leq j}[\underline{s_1 \dots s_j}, :]^\top\right)\\
&= \sum_{s_1, ..., s_{k-1}}
\frac{1}{R_j} \mathrm{Tr} \br{A_{\leq j}[\underline{s_1 \dots s_j}, :]^\top \cdot A_{\leq j}[\underline{s_1 \dots s_j}, :]} \\
&= \frac{1}{R_j} \sum_{s_1, ..., s_{k-1}}
\mathrm{Tr} \br{\scr A_j \br{:, s_j, :}^\top \cdot ... \cdot
\scr A_1 \br{:, s_1, :}^\top \cdot 
\scr A_1 \br{:, s_1, :} \cdot ... \cdot \scr A_j \br{:, s_j, :}} \\
&= \frac{1}{R_j} \sum_{s_2, ..., s_{k-1}}
\mathrm{Tr} \br{\scr A_j \br{:, s_j, :}^\top \cdot ... \cdot
\paren{\sum_{s_1} \scr A_1 \br{:, s_1, :}^\top \cdot 
\scr A_1 \br{:, s_1, :}} \cdot ... \cdot \scr A_j \br{:, s_j, :}} \\
&= \frac{1}{R_j} \sum_{s_2, ..., s_{k-1}}
\mathrm{Tr} \br{\scr A_j \br{:, s_j, :}^\top \cdot ... \cdot
\scr A_2 \br{:, s_2, :}^\top \cdot I \cdot 
\scr A_2 \br{:, s_2, :} \cdot ... \cdot \scr A_j \br{:, s_j, :}}.
\end{aligned}
\end{equation}
In the expressions above, the summation over each variable
$s_t$, $1 \leq t \leq k$, is taken over the range $[I_t]$.
The first step follows by marginalizing over random variables
$\hat s_1, ..., \hat s_{k-1}$. The second
step follows from Equation \eqref{eq:hat_s_defn}. The
third step rewrites an inner product of two vectors as the
trace of their outer product. The fourth step follows
from the definition of $A_{\leq j}$. The fifth step follows
from the linearity of the trace by moving the summation
over $s_1$ into the product expression. The last step
follows from the definition of the left-orthonormality property
on $\scr A_1$; that is, $\sum_{s_1} 
\scr A_1 \br{:, s_1, :}^\top \cdot \scr A \br{:, s_1, :} =
A^{L \top}_1 A^{L}_1 = I$. By successively moving summation 
operators into the product expression to repeat the last step
(exploiting the left-orthonormality of each core 
in the process), we find
\begin{equation}
\begin{aligned}
p(&\hat s_k = s_k\ \vert\ \hat s_{>k} = s_{>k}) \\
&= \frac{1}{R_j} 
\mathrm{Tr} \br{\scr A_j \br{:, s_j, :}^\top \cdot ... \cdot
\scr A_k \br{:, s_k, :}^\top \cdot 
\scr A_k \br{:, s_k, :} \cdot ... \cdot \scr A_j \br{:, s_j, :}} \\
&= \frac{1}{R_j} 
\mathrm{Tr} \br{H_{>k}^\top \cdot
\scr A_k \br{:, s_k, :}^\top \cdot 
\scr A_k \br{:, s_k, :} \cdot H_{>k}}, \\
\end{aligned}
\end{equation}
where the last line follows from the definition of $H_{>k}$.
\end{proof}

\subsection{Proof of Lemma \ref{lemma:equivalent_conditional_distribution}}
\label{appendix:equivalent_lemma_proof}

\begin{proof}[\unskip\nopunct]
We write
\begin{equation}
\begin{aligned}
p(\hat s_k = s_k\ \vert\ \hat s_{>k} = s_{>k})
&= \frac{1}{R_j} 
\mathrm{Tr} \br{H_{>k}^\top \cdot
\scr A_k \br{:, s_k, :}^\top 
\scr A_k \br{:, s_k, :} \cdot H_{>k}} \\
&= \frac{1}{R_j} 
\sum_{r=1}^{R_j} \paren{e_r^\top \cdot H_{>k}^\top \cdot
\scr A_k \br{:, s_k, :}^\top \cdot
\scr A_k \br{:, s_k, :} \cdot H_{>k} \cdot e_r} \\
&= \frac{1}{R_j} 
\sum_{r=1}^{R_j} \paren{h_{>k}^\top \cdot
\scr A_k \br{:, s_k, :}^\top \cdot 
\scr A_k \br{:, s_k, :} \cdot h_{>k}} \\
&= \frac{1}{R_j} 
\sum_{r=1}^{R_j} 
p(\hat t_k = t_k\ \vert\ \hat t_{>k} = t_{>k}, \hat r = r). 
\end{aligned}
\end{equation}
The first step follows from Lemma \ref{lemma:conditional_distribution}.
The second step follows from the definition of the trace.
The third step follows from the definitions of $h_{>k}$ and 
$H_{>k}$. The fourth step follows from the definition of
the variables $\hat t_1, ..., \hat t_j$. Now observe
that $p(\hat r = r) = 1 / R_j$ for $1 \leq r \leq R_j$, 
so we can write

\begin{equation}
\begin{aligned}    
p(\hat s_k = s_k\ \vert\ \hat s_{>k} = s_{>k})
&= \sum_{r=1}^{R_j} 
p(\hat t_k = t_k\ \vert\ \hat t_{>k} = t_{>k}, \hat r = r) p(\hat r = r) \\
&=p(\hat t_k = t_k\ \vert\ \hat t_{>k} = t_{>k}),
\end{aligned}
\end{equation}
which completes the proof.
\end{proof}

\subsection{Efficient Sampling Data Structure}
\label{appendix:efficient_sampling_ds}
\begin{proof}[\unskip\nopunct]
Lemma \ref{lemma:efficient_row_sampler} first
appeared as Lemma 3.2 in the original work
by \citet{bharadwaj2023fast}. We state a 
condensed form of the original claim below: 

\begin{lemma}[\citet{bharadwaj2023fast}, Original]
Given $U \in \RR^{M \times R}$, 
$Y \in \RR^{R \times R}$ with $Y$ p.s.d., there 
exists a data structure parameterized by
positive integer $F$ that requires $O(MR^2)$
time to construct and additional space space $O(R^2 \lceil M / F \rceil)$. After construction,
the data structure can draw a sample from the
un-ndistribution defined elementwise by
$$q_{h, U, Y}\br{s} := C^{-1} U\br{s, :} 
\paren{Y \circledast h h^\top}  U\br{s, :}^\top$$
in time $O(R^2 \log \lceil M / F \rceil + FR^2)$. When $Y$ is a rank-1 matrix, the runtime drops 
to $O(R^2 \log \lceil M / F \rceil + FR)$.
\end{lemma}
In the lemma above, $C$ is an appropriate
normalization constant. The datas structure
that this lemma describes relies on a binary tree
data structure that is truncated to 
$\log \lceil I / F \rceil$ levels. To draw
a sample, the data structure executes
a random walk that requires $O(R^2)$ work at
each internal node and some additional
computation at the leaf.

To prove our adapted lemma, 
take $Y = \br{1}$, a matrix of all ones that 
is rank-1, and set $F = R$. Then 
$$q_{h, U, Y}\br{s} = C^{-1}U\br{s, :} 
\paren{h h^\top}  U\br{s, :}^\top =
C^{-1} (U\br{s, :} \cdot h)^2$$
This is the target probability distribution
of Lemma \ref{lemma:efficient_row_sampler},
and the runtime to draw each sample is
$O(R^2 \log(M / R) + R^2) = O(R^2 \log (M / R))$.
The choice $F = R$ also induces space usage
$O(MR)$, linear in the size of the input. Our 
modified claim follows.
\end{proof}

\subsection{Proof of Theorem \ref{thm:main_result}}
\label{sec:main_proof}
\begin{proof}[\unskip\nopunct]
We provide a short end-to-end proof that shows 
that Algorithms 
\ref{alg:chain_sampler_construction} and
\ref{alg:chain_sampling} correctly draw samples
from $A_{\leq j}$ (the matricization of the
left-orthogonal core chain) according to the
distribution of its squared row norms while meeting 
the runtime and space guarantees of Theorem
\ref{thm:main_result}.

\textbf{Construction Complexity}: The cost 
of Algorithm \ref{alg:chain_sampler_construction}
follows from \ref{lemma:efficient_row_sampler} with
$M = I R_{k-1}$, the row count of $A^L_k$ for
$1 \leq k \leq j$. Using this lemma, construction of 
each sampling data structure $Z_k$ requires 
time $O(I_k R_{k-1} R_k^2)$. The space required by
sampler $Z_k$ is $O(I_k R_{k-1} R_k)$; summing over
all indices $k$ gives the construction claim in
Theorem \ref{thm:main_result}.
\newline

\textbf{Sampling Complexity: } The complexity to
draw samples in Algorithm \ref{alg:chain_sampling}
is dominated by calls to the RowSample procedure,
which as discussed in Section \ref{efficient-tt-sampling} is
$O(R_k^2 \log(I_k R_{k-1} / R_{k}))$
Summing the complexity over indices
$1 \leq k < j$ yields the cost claimed by
Theorem \ref{thm:main_result} to draw a single
sample. The complexity of calling the RowSample procedure
repeatedly dominates the complexity to update the
history vector $h$ over all loop iterations, 
which is $O\paren{\sum_{k=1}^j R_{k-1} R_k}$ for each sample.
\newline

\textbf{Correctness: } Our task is to show that Algorithm
\ref{alg:chain_sampling} each sample $t_d$, $1 \leq d \leq J$,
is a multi-index that follows the squared row norm 
distribution on the rows of $A_{\leq j}$. To do this, 
we rely on lemmas proven earlier. For each
sample, the variable $\hat r$ is a uniform random draw
from $\br{R_j}$, and $h$ is initialized to the
corresponding basis vector. By Equation 
\eqref{eq:consecutive_entries} and Lemma 
\ref{lemma:efficient_row_sampler}, Line 5 from
Algorithm \ref{alg:chain_sampling} draws each index
$\hat t_k$ correctly according to the probability
distribution specified by Equation \eqref{eq:hat_t_defn}.
The history vector is updated by Line 6 of the algorithm
so that subsequent draws past iteration $k$ of the loop are
also drawn correctly according to Equation 
\eqref{eq:hat_t_defn}. Lemma 
\ref{lemma:equivalent_conditional_distribution} (relying
on Lemma \ref{lemma:conditional_distribution}) 
shows that
the multi-index $\underline{\hat t_1 \dots \hat t_j}$ drawn
according to Equation \eqref{eq:hat_t_defn} follows the same
distribution as $\underline{\hat s_1 \dots \hat s_j}$, which was 
defined to follow the squared norm distribution on the
rows of $A_{\leq j}$. This completes the proof.
\end{proof}
\subsection{Proof of Corollary~\ref{cor:rtt-als-alg}}
\label{sec:corollary-4.4}
Since $A^{\neq j}\in\RR^{\prod_{k\neq j}^NI_k\times R_{j-1}R_{j}}$ and $X_{(j)}\in\RR^{\prod_{k\neq j}^NI_k\times I_j}$, we draw $\tilde{O}(R^2/\varepsilon\delta)$ samples to achieve the error bound $(1+\varepsilon)$ with probability $(1-\delta)$ for each least squares solve in the down-sampled problem~(\ref{thm:relative-error gaurantee}). By Theorem~\ref{thm:main_result}, the complexity of drawing $J$ samples with our data structure
is 
$$O\paren{\sum_{k\neq j}J\log I_kR^2} = \tilde{O}\paren{\sum_{k\neq j}R^4/(\varepsilon\delta) \log I_k}$$
where we suppose that $R_1=R_2=\dots=R_{N-1}$ and $I_1=\dots=I_N$. The cost of sampling a corresponding subset of $X_{(j)}$ is $O(JI_j) = \tilde{O}\paren {R^2/(\varepsilon\delta) I_j}$. Solving the downsampled least squares problem also costs $O(JR^2 I_j) = \tilde{O}\paren{I_j R^4/(\varepsilon\delta)}$. Summing them all together for $1 \leq j\leq N$ gives
\begin{align}
    &\Tilde{O}\paren{1/\varepsilon\delta\paren{\sum_{j=1}^N\paren{\sum_{k\neq j}R^4\log I_k} + R^4I_j}}\nonumber\\
    &=
    \Tilde{O}\paren{R^4/\varepsilon\delta \cdot \sum_{j=1}^N (N-1)\log I_j + I_j}\nonumber\\
    &=
  \Tilde{O}\paren{R^4/\varepsilon\delta \cdot \sum_{j=1}^N N\log I_j + I_j}\nonumber
\end{align}
where we wrote the last equation considering the fact that $N$ dominates $(N-1)$.
\newpage

\section{Detail about Datasets \& Experiments}\label{app:datasets}

\subsection{Datasets}

For the real dense datasets experiment, we  truncated and reshaped the original data tensors in to the fourth order tensors as follows.  

\begin{itemize}
    \item \textbf{Pavia University dataset} original tensor is of size $(610,340,103)$. We truncate it to $(600,320,100)$ and permute the modes to have the size $(100,600,320)$ tensor and reshaped into a tensor of size $(100,320,24,25)$. It is downloaded from
    
\url{http://lesun.weebly.com/hyperspectral-data-set.html}
\item \textbf{Tabby Cat dataset} is reshaped to a tensor of size $(286,720,40,32)$. The video is in color and converted to grayscale by averaging the three color channels. It is downloaded from

\url{https://www.pexels.com/video/video-of-a-tabby-cat-854982/}. 

    \item \textbf{The MNIST dataset} is reshaped into a tensor of size $(280,600,28,10)$ and is downloaded from
    
    \url{https://www.kaggle.com/datasets/hojjatk/mnist-dataset}
    \item \textbf{The Washington DC Mall dataset} is truncated to $(1280,306,190)$ and reshaped into  a tensor of size $(1280,306,10,19)$ and  is downloaded from
    
\url{https://engineering.purdue.edu/˜biehl/MultiSpec/hyperspectral.html.}
\end{itemize}

\subsection{Computing Resources}
All experiments are executed on CPU nodes of institutional clusters. The dense data experiments are run on nodes with 4 CPUs and 16GB of RAM.

\end{document}